\documentclass[onecolumn, draft, 12pt]{IEEEtran}
\usepackage{amsmath,cases}
\usepackage{amsthm}
\usepackage{amssymb}
\usepackage{cite}
\usepackage{amsfonts}
\usepackage{enumerate}
\usepackage[final]{graphicx}
\usepackage{multirow}
\usepackage{caption}

\newcommand{\powcnst}{\ensuremath{P_{\rm T}}}

\newcommand{\sigv}{\ensuremath{\sigma_v^2}}

\newcommand{\w}{\ensuremath{\omega}}

\newcommand{\tphii}{\ensuremath{{\widetilde \phi}_i}}

\newtheorem{thm}{Theorem}

\begin{document}
\title{Distributed Detection over Gaussian Multiple Access Channels with Constant Modulus Signaling}
\author{Cihan Tepedelenlio\u{g}lu, \emph{Member, IEEE}, Sivaraman Dasarathan
\thanks{The authors are with the School of Electrical, Computer, and Energy Engineering, Arizona
State University, Tempe, AZ 85287, USA. (Email:
\{cihan, sdasarat\}@asu.edu).} } \maketitle

\begin{abstract}
A distributed detection scheme where the sensors transmit with constant modulus signals over a Gaussian multiple access channel is considered. The deflection coefficient of the proposed scheme is shown to depend on the characteristic function of the sensing noise and the error exponent for the system is derived using large deviation theory. Optimization of the deflection coefficient and error exponent are considered with respect to a transmission phase parameter for a variety of sensing noise distributions including impulsive ones. The proposed scheme is also favorably compared with existing amplify-and-forward and detect-and-forward schemes. The effect of fading is shown to be detrimental to the detection performance through a reduction in the deflection coefficient depending on the fading statistics. Simulations corroborate that the deflection coefficient and error exponent can be effectively used to optimize the error probability for a wide variety of sensing noise distributions.
\end{abstract}

\section{Introduction} \label{sec:intro}
In inference-based wireless sensor networks, low-power sensors with limited battery and peak power capabilities transmit their observations to a fusion center (FC) for detection of events or estimation of parameters. For distributed detection, much of the literature has focused on the parallel topology where each sensor uses a dedicated channel to transmit to a fusion center. Multiple access channels offer bandwidth efficiency since the sensors transmit over the same time/frequency slot. The sensors may perform local detection through quantization in which case the decision is encoded into a specific waveform to be sent to the FC. Instead, sensed information may be sent to the FC using analog modulation which transmits the unquantized data by appropriately pulse shaping and amplitude or phase modulating to consume finite bandwidth.

In \cite{tolga}, the distributed detection over a multiple access channel is studied where arbitrary number of quantization levels at the local sensors are allowed, and transmission from the sensors to the fusion center is subject to both noise and inter-channel interference. References \cite{lidai, Sayeed_Type, Tong_Asymp_fading, Tong_Random} discuss distributed detection over Gaussian multiple access channels. In \cite{lidai}, detection of a deterministic signal in correlated Gaussian noise and detection of a first-order autoregressive signal in independent Gaussian noise are studied using an amplify-and-forward scheme where the performance of different fusion rules is analyzed.  In \cite{Sayeed_Type}, a type-based multiple access scheme is considered in which the local mapping rule encodes a waveform according to the type \cite[pp. 347]{cover} of the sensor observation and its performance under both the per-sensor and total power constraints is investigated. This scheme is extended to the case of fading between the sensors and the FC in \cite{Tong_Asymp_fading} and its performance is analyzed using large deviation theory. In the presence of non-coherent fading over a Gaussian multi-access channel, type-based random access is proposed and analyzed in \cite{Tong_Random}. In \cite{Coyle2007}, the optimal distributed detection scheme in a clustered multi-hop sensor network is considered where a large number of distributed sensor nodes quantize their observations to make local hard decisions about an event. The optimal decision rule at the cluster head is shown to be a threshold test on the weighted sum of the local decisions and its performance is analysed.

Two schemes called modified amplify-and-forward (MAF) and the modified detect-and-forward (MDF) are developed in \cite{Evans_DD_MA} which generalize and outperform the classic amplify-and-forward (AF) and detect-and-forward (DF) approaches to distributed detection. It is shown that MAF outperforms MDF when the number of sensors is large and the opposite conclusion is true when the number of sensors is smaller. For the MDF scheme with identical sensors, the optimal decision rule is proved to be a threshold test in  \cite{Evans_Optimal}. Decision fusion with a non-coherent fading Gaussian multiple access channel is considered in \cite{Evans_Fading} where the optimal fusion rule is shown to be a threshold test on the received signal power and on-off keying is proved to be the optimal modulation scheme. A distributed detection system where sensors transmit their observations over a fading Gaussian multiple-access channel to a FC with multiple antennas using amplify-and-forward is studied in \cite{banavar}. In all these cases, the sensing noise distribution is assumed to be Gaussian. Even though the Gaussian assumption is widely used, sensor networks which operate in adverse conditions require detectors which are robust to non-Gaussian scenarios. Moreover, in the literature there has been little emphasis on distributed schemes with the desirable feature of using constant modulus signals with fixed instantaneous power.

A distributed estimation scheme where the sensor transmissions have constant modulus signals is considered in \cite{tepadarsh}. Distributed estimation in a bandwidth-constrained sensor network with a noisy channel is investigated in \cite{Aysal2008} and distributed estimation of a vector signal in a sensor network with power and bandwidth constraints is studied in \cite{Xiao2008}. The estimator proposed in \cite{tepadarsh} is shown to be strongly consistent for any sensing noise distribution in the iid case. Inspired by the robustness of this estimation scheme, in this work, a distributed detection scheme where the sensors transmit with constant modulus signals over a Gaussian multiple access channel is proposed for a binary hypothesis testing problem. The sensors transmit with constant modulus transmissions whose phase is linear with the sensed data. The output-signal-to-noise-ratio, also called as the deflection coefficient (DC) of the system, is derived and expressed in terms of the characteristic function (CF) of the sensing noise. The optimization of the DC with respect to the transmit phase parameter is considered for different distributions on the sensing noise including impulsive ones. The error exponent  is also derived and shown to depend on the CF of the sensing noise. It is shown that both the DC and the error exponent can be used as accurate predictors of the phase parameter that minimizes the detection error rate. The proposed detector is favorably compared with MAF and the MDF schemes developed in \cite{Evans_DD_MA,Evans_Optimal} for the Gaussian sensing noise and its robustness in the presence of other sensing noise distributions is highlighted. The effect of fading between the sensors and the fusion center is shown to be detrimental to the detection performance through a reduction in the DC depending on the fading statistics. Different than \cite{tepadarsh} where the asymptotic variance of an estimator is analyzed, the emphasis herein is on derivation, analysis, and optimization of detection-theoretic metrics such as the DC and error exponent. Our aim in this paper is to develop a distributed detection scheme where the instantaneous transmit power is not influenced by possibly unbounded sensor measurement noise. 

The paper is organized as follows. In Section \ref{sec: model}, the system model is described with per-sensor power constraint and total power constraint. In Section \ref{sec: Detection_problem}, the detection problem is described  and a linear detector is proposed. The probability of error performance of the detector is analyzed in Section \ref{sec: analysis_prob_error}. The DC is defined and its optimization for several cases is studied in Section \ref{sec: analysis_n_opt_D_w}. The presence of fading between the sensors and the fusion center is discussed in Section  \ref{sec:fading_channels}. The error exponent of the proposed detector is analyzed in Section \ref{sec:error_exponent}. Non-Gaussian channel noises are discussed in Section \ref{sec: nonGaussianChannel}. Simulation results are provided in Section \ref{sec: simulations} which support the theoretical results. Finally, the concluding remarks are presented in Section \ref{Conclusions}.

\section{System Model} \label{sec: model}
Consider a binary hypothesis testing problem with two hypotheses $H_0$, $H_1$ where $P_0$, $P_1$ are their respective prior probabilities. Let the sensed signal at the $i^{th}$ sensor be,
\begin{equation}
\label{eq:system_model}
x_i =
\begin{cases} 
\theta + n_{i}& \mathrm{under \:} H_{1}\\
n_{i}  & \mathrm{under \:} H_{0}\end{cases}
\end{equation}
$i=1, \ldots, L$, $\theta>0$ \footnote{the proposed scheme will work without any difference for $\theta <0$ due to symmetry if we substitute $-\theta$ in the place of $\theta$ in all the equations.} is a known parameter whose presence or absence has to be detected, $L$ is the total number of sensors in the system, and $n_i$ is the noise sample at the $i^{th}$ sensor. The sensing noise samples are independent, have zero median and an absolutely continuous distribution but they need not be identically distributed or have any finite moments. We consider a setting where the $i^{th}$ sensor transmits its measurement using a constant modulus signal $\sqrt{\rho}e^{j \omega x_{i}}$ over a Gaussian multiple access channel so that the received signal at the FC is given by
\begin{equation}
\label{eq:recd_signal}
y_L = \sqrt{\rho}\displaystyle\sum_{i=1}^{L} e^{j \omega x_{i}} + v
\end{equation}
where $\rho$ is the power at each sensor, $\omega>0$ is a design parameter to be optimized and $v \sim  \mathcal{CN}(0, \sigma^{2}_{v}) $ is the additive channel noise. We consider two types of power constraints: Per-sensor power constraint and total power constraint. In the former case, each sensor has a fixed power $\rho$ so that the total power $P_{\rm T}=\rho L$, and as $L \rightarrow \infty$, $P_{\rm T} \rightarrow \infty$; in the later case, the total power $P_{\rm T}$ is fixed for the entire system and does not depend on $L$, so that the per-sensor power $\rho=P_{\rm T}/L \rightarrow 0$ as $L \rightarrow \infty$. 

\section{The Detection Problem} \label{sec: Detection_problem}
The received signal $y_L$ under the total power constraint can be written as
\begin{equation}
\label{eq:recd_signal_PT}
y_L = \sqrt{\frac{P_{\rm T}}{L}}\displaystyle\sum_{i=1}^{L} e^{j \omega x_{i}} + v .
\end{equation}
We assume throughout that $P_0=P_1=0.5$ for convenience even though other choices can be easily incorporated. With the received signal in (\ref{eq:recd_signal_PT}), the FC has to decide which hypothesis is true. It is well known that the optimal fusion rule under the Bayesian formulation is given by:
\begin{equation}
\label{eq:opt_fusion_rule}
\frac{f(y_L|H_1)}{f(y_L|H_0)} \overset{H_{1}}{\underset{H_{0}}{\gtrless}} \frac{P_0}{P_1}=1
\end{equation}
where $f(y_L|H_i)$, is the conditional probability density function of $y_L$ when $H_i$ is true. 
The equation (\ref{eq:recd_signal_PT}) can be rewritten as follows:
\begin{equation}
\label{eq:recd_signal_PT_cos_sin}
\nonumber
y_L = \sqrt{\frac{P_{\rm T}}{L}} \left ( \displaystyle\sum_{i=1}^{L} \cos (\omega x_{i}) \right ) + j \sqrt{\frac{P_{\rm T}}{L}} \left ( \displaystyle\sum_{i=1}^{L} \sin (\omega x_{i}) \right ) + v .
\end{equation}
Since there are $L$ terms in the first summation involving the cosine function, we need to do $L$ fold convolutions with the PDFs of $\cos (\omega x_{i})$ and another set of $L$ fold convolutions with the PDFs of $\sin (\omega x_{i})$. Then we need to find the joint distribution of the PDFs obtained thus for the cosine and sine counterparts. This joint PDF will need to be convolved with the PDF of $v$. It is not possible to obtain a closed form expression for these $(2L+1)$ fold convolutions. Hence, $f(y_L|H_i)$ is not tractable. Therefore, we consider the following linear detector which is argued next to be optimal for large $L$:
\begin{equation}
\label{eq:detector_gaussian}
\Re [y_L e^{-j\omega \theta }] - \Re [y_L] \overset{H_{1}}{\underset{H_{0}}{\gtrless}} 0 \;,
\end{equation} where we define $\Re[y]$ as the real part, and $\Im[y]$ as the imaginary part of $y$. Note that the detector in (\ref{eq:detector_gaussian}) would be optimal if $y_L$ were Gaussian. Clearly due to central limit theorem $y_L$ in \eqref{eq:recd_signal_PT} is asymptotically Gaussian, which indicates that (\ref{eq:detector_gaussian}) approximates  \eqref{eq:opt_fusion_rule}  for large $L$. With the Gaussian assumption, the variances of $y_L$ in \eqref{eq:recd_signal_PT} under the two hypotheses are the same and given by Var$(y_L|H_0)=$Var$(y_L|H_1)= [P_{\rm T} (1-\varphi_n^{2} (\omega)) + \sigma_{v}^{2}]$, where $\varphi_n(\omega)$ is the characteristic function of $n_i$. Hence, the optimal likelihood ratio simplifies to the detector in (\ref{eq:detector_gaussian}) which is linear in $y_L$, when $y_L$ is assumed Gaussian which holds for large $L$. However as will be seen in Section \ref{sec: analysis_prob_error}, we do not assume that $y_L$ is Gaussian for any fixed $L$ when we analyze the performance of the detector in (\ref{eq:detector_gaussian}) or in finding the associated error exponent in Section \ref{sec:error_exponent}. We proceed by expressing the probability of error.

\section{Probability of Error} \label{sec: analysis_prob_error}
The detector in (\ref{eq:detector_gaussian}) depends on the design parameter $\omega$ and this means that the probability of error will in turn depend on $\omega$. Let $P_{\rm e}(\omega)$ be the probability of error at the FC:
\begin{equation}
\label{eq:prob_error_bayesian}
P_{\rm e}(\omega) = \frac{1}{2} \Pr \left [{\rm error}|H_0 \right ] + \frac{1}{2} \Pr \left [{\rm error}|H_1 \right ] = \Pr \left [{\rm error}|H_0 \right ]
\end{equation}
where $\Pr \left [{\rm error}|H_i \right ]$ is the error probability when $H_i$, $i \in \lbrace 0,1 \rbrace$, is true and the last equality holds due to symmetry between the two hypotheses which is explained as follows. From the detection rule \eqref{eq:detector_gaussian}, the probability of error under $H_0$ is given by
\begin{equation}
\label{eq:prob_error_bayesian_tmp}
\Pr \left [{\rm error}|H_0 \right ]= \Pr \left [ \Re [y_L]  < \Re [y_L e^{-j\omega \theta }] | H_0 \right ] \;,
\end{equation}
where the received signal in \eqref{eq:recd_signal_PT} under $H_0$ is given by
\begin{equation}
\label{eq:y_L_alt}
y_L = \left( \sqrt{\frac{P_{\rm T}}{L}} \displaystyle\sum_{i=1}^{L} \cos(\omega n_{i}) + \Re [v] \right ) + j \left( \sqrt{\frac{P_{\rm T}}{L}}\displaystyle\sum_{i=1}^{L} \sin(\omega n_{i}) + \Im [v] \right ).
\end{equation}
Substituting \eqref{eq:y_L_alt} for $y_L$ in \eqref{eq:prob_error_bayesian_tmp} and doing some algebraic simplifications we get,
\begin{equation}
\label{eq:prob_error_bayesian2}
\Pr \left [{\rm error}|H_0 \right ] = \Pr\left[ \underbrace{\displaystyle\sum_{i=1}^{L} 2 \sin \left (\frac{\omega \theta}{2} \right ) \cos \left(\omega n_{i} - \frac{\omega \theta}{2}+\frac{\pi}{2}\right ) + \sqrt{\frac{L}{P_{\rm T}}} v^{\rm T} }_{Z_L(\omega):=} <  0 \right] \;,
\end{equation}
where $v^{\rm T}:=\Re [v](1-\cos(\omega \theta)) - \Im [v]\sin(\omega \theta)$. Similarly, $\Pr \left [{\rm error}|H_1 \right ]$ is same as that of \eqref{eq:prob_error_bayesian2} except the argument of the cosine function is replaced by $(\omega n_{i}+\omega \theta/2 - \pi /2)$. To see the symmetry between the two hypotheses asserted in \eqref{eq:prob_error_bayesian}, let $\zeta:=(\omega \theta/2 - \pi /2)$ for convenience, so that $\cos(\omega n_{i} \mp \zeta)=[\cos(\omega n_{i}) \cos \zeta + \sin( \pm \omega n_{i}) \sin \zeta]$. Since $n_i$ is symmetric, $\omega n_i$ and $-\omega n_i$ have the same distribution which implies that the random variables $\cos( \omega n_i -\zeta)$ and $\cos( \omega n_i +\zeta)$ have the same distribution establishing that $\Pr \left [{\rm error}|H_1 \right ]=\Pr \left [{\rm error}|H_0 \right ]$. Therefore, the probability of error in \eqref{eq:prob_error_bayesian} is given by \eqref{eq:prob_error_bayesian2}. We are interested in using \eqref{eq:prob_error_bayesian2} to find the $\omega$ that minimizes the probability of error at FC. Since $P_{\rm e}(\omega)$ is not straightforward to evaluate, we optimize two surrogate metrics to select $\omega$. These are the error exponent and the DC. The error exponent is an asymptotic measure of how fast the $P_{\rm e}(\omega)$ decreases as $L \rightarrow \infty$, and is specific to the detector used in \eqref{eq:detector_gaussian} and will be considered in Section \ref{sec:error_exponent}. The DC, on the other hand, is specific to the model in \eqref{eq:recd_signal_PT}, and does not depend on any detector.

\section{Deflection Coefficient and its Optimization} \label{sec: analysis_n_opt_D_w}
We will now define and use the deflection coefficient which reflects the output-signal-to-noise-ratio and widely used in optimizing detectors \cite{picinbono, varshney, poor, kassam}. The DC is mathematically defined as,
\begin{equation}
\label{eq:defl_coef_def}
D(\omega):=\frac{1}{L} \frac{|{\rm E}[y_L|H_1] - {\rm E}[y_L|H_0] |^{2}}{{\rm var}[y_L|H_0]} .
\end{equation}
By calculating the expectations in \eqref{eq:defl_coef_def}, it can be easily verified that the DC for the signal model in  (\ref{eq:recd_signal}) is given by: 
\begin{equation}
\label{eq:defl_coef_PT}
D(\omega)= \frac{ 2 \varphi_n^{2} (\omega)[1 - \cos (\omega \theta)]} {\left [1-\varphi_n^{2} (\omega) + \frac{\sigma_{v}^{2}}{P_{\rm T}}\right]}
\end{equation}
where $\varphi_n(\omega)={\rm E}[e^{j \omega n_{i}}]$ is the CF of $n_{i}$. The CF $\varphi_n(\omega)$ does not depend on the sensor index $i$, since we will be initially assuming that $n_i$ are iid. We will consider the non-identically distributed case in Section \ref{sec: Non-homogeneous}. Note that $D(\omega) \geq 0$ and that $\varphi_n(\omega)$ is real-valued since $n_i$ is a symmetric random variable. Moreover, $\varphi_n(\omega)=\varphi_n(-\omega)$ so that $D(\omega)=D(-\omega)$ which justifies why we will focus on $\omega >0$ throughout. The factor $(1/L)$ introduced in (\ref{eq:defl_coef_def}) does not appear in conventional definitions of the DC but included here for simplicity since it does not affect the optimal $\omega$.

\subsection{Optimizing $D(\omega)$} \label{sec: optimize_D_w}
We are now interested in finding $\omega$ by optimizing $D(\omega)$:
\begin{equation}
\label{eq:omega_opt}
\omega^{*}:= {\arg\max_{\omega > 0}}  \; D(\omega) .
\end{equation}
Since $\varphi_n(\omega)\leq 1$, when $\sigma_{v}^{2}>0$, $D(\omega)$ is bounded, and achieves its smallest value of $D(\omega)=0$ as $\omega \rightarrow 0$. On the other hand, as $\omega \rightarrow \infty$, we have $\lim_{\omega \rightarrow \infty} D(\omega)=0$. This implies that the maximum in \eqref{eq:omega_opt} cannot be achieved by $\omega=0$ or $\omega=\infty$ and establishes that there must be a finite $\omega^{*} \in (0, \infty)$ which attains the maximum in \eqref{eq:omega_opt}.

In what follows, we will further characterize $\omega^{*}$ by assuming that $\varphi_n(\omega)> 0$ and $\varphi_n^{'}(\omega)< 0$ for all $\omega > 0$. Many distributions including the Laplace, Gaussian and Cauchy have CFs that satisfy this assumption. Indeed all symmetric alpha-stable distributions \cite[pp. 20]{alpha_stable} of which the latter two is a special case, satisfy this assumption. We now have the following theorem which restricts $\omega^{*}$ in \eqref{eq:omega_opt} to a finite interval.
\begin{thm} \label{thm1}
If $\varphi_n(\omega)$ is decreasing and differentiable over $\omega>0$, then $\omega^{*} \in (0, \pi/\theta)$.
\end{thm}
\begin{IEEEproof}
First, note that $\varphi_n(\omega) \geq 0$ which is implied by the assumption that $\varphi_n(\omega)$ is decreasing and that $\varphi_n(\omega) \rightarrow 0$ as $\omega \rightarrow \infty$. Let $D(\omega)=C(\omega) [1-\cos(\omega \theta)]$ with $C(\omega):=2 \varphi_n^{2} (\omega) / [1-\varphi_n^{2} (\omega) + \sigma_{v}^{2} / P_{\rm T} ]$ for brevity. Since $\varphi_n(\omega)$ is decreasing on $\omega>0$ and $\varphi_n(\omega) \geq 0$, $C(\omega)$ is also decreasing. Because $[1-\cos(\omega \theta)]$ is periodic in $\omega$ with period  $2\pi / \theta$,
\begin{equation}
\label{eq:theorem1_eq1}
D \left ( \omega + \frac{2 \pi} {\theta}  \right ) = [1-\cos(\omega \theta)] C \left ( \omega + \frac{2 \pi} {\theta}  \right ) < [1-\cos(\omega \theta)] C( \omega) =D (\omega).
\end{equation}
Noticing that $D ( 2 \pi / \theta)=0$ which rules out $\omega^{*} = 2\pi/\theta$, we have $\omega^{*} \in (0, 2\pi/\theta)$. To further reduce the range of $\omega^{*}$ by half, consider the fact that $D(0)=D ( 2 \pi / \theta)=0$, which combined with $D(\omega)>0$ for $\omega \in (0, 2\pi/\theta)$ implies that $\omega^{*} \in (0, 2\pi/\theta)$ satisfies $D^{'}(\omega^{*})=0$. Writing $D^{'}(\omega^{*})=0$ we obtain:
\begin{equation}
\label{eq:theorem1_eq2}
\frac{[\theta \sin(\omega^{*} \theta)]}{[\cos(\omega^{*} \theta)-1]} = \frac{C^{'}(\omega^{*})}{C(\omega^{*})} .
\end{equation}
Since $C(\omega)>0$ is decreasing, the right hand side (rhs) of \eqref{eq:theorem1_eq2} is negative and it follows that $\omega^{*} \in (0, \pi/\theta)$ as required.
\end{IEEEproof}

By the definition of $\omega^{*}$, it is clearly a function of $\theta$. We showed in Theorem \ref{thm1} that $0 < \omega^{*} < \pi/\theta$ if $\varphi_n^{'}(\omega)< 0$ for $\omega>0$. Note that when $\omega=0$, there is no phase modulation done, and what is transmitted is a constant signal which actually contains no information about $x_i$. Therefore the boundary value $\omega=0$ is not a valid choice. When $\omega=\pi/\theta$, the detector in \eqref{eq:detector_gaussian} actually simplifies to: $\Re[y_L] \overset{H_0}{\underset{H_1}{\gtrless}} 0$. While $\omega=\pi/\theta$ is a valid choice, it is optimal only when $\theta$ is large as will be proved in Theorem 2. We now investigate the behavior of $\omega^{*}$ when $\theta$ is large without assuming anything about $\varphi_n(\omega)$ except the absolute continuity of its distribution, and show that $\omega^{*}\approx \pi/\theta$ for large $\theta$ in the sense that $\omega^{*} \theta \rightarrow \pi$, as $\theta \rightarrow \infty $.
\begin{thm} \label{thm2}
If $\sigma_{v}^{2}>0$, and $n_i$ are iid and have absolutely continuous distributions,
\begin{equation}
\lim_{\theta \rightarrow \infty} \omega^{*} \theta = \pi .
\end{equation}
\end{thm}
\begin{IEEEproof}
We have
\begin{equation}
\label{eq:theorem2_eq1}
D \left (\frac{\pi}{\theta} \right ) \leq D(\omega^{*}) \leq \sup_{\omega>0} [1-\cos(\omega \theta)] \; \sup_{\omega>0} C(\omega)=\frac{4 P_{\rm T}}{\sigma_{v}^{2}} \;,
\end{equation}
where the first inequality is because $\omega^{*}$ maximizes $D(\omega)$, and the second inequality follows from $D(\omega)=C(\omega) [1-\cos(\omega \theta)]$. Recalling that $\lim_{\omega \rightarrow 0} \varphi_n(\omega)=1$ we take the limit as $\theta \rightarrow \infty$ in \eqref{eq:defl_coef_PT} and obtain $\lim_{\theta \rightarrow \infty} D(\pi /\theta)= 4 P_{\rm T} / \sigma_{v}^{2}$, which using \eqref{eq:theorem2_eq1} shows that $\lim_{\theta \rightarrow \infty} D(\omega^{*})= 4 P_{\rm T} / \sigma_{v}^{2}$. Since $\varphi_n(0) > \varphi_n(\omega)$ and because $D(\omega)$ is an increasing function of $\varphi_n^{2}(\omega)$, from  \eqref{eq:defl_coef_PT}  it is clear that the only way $\lim_{\theta \rightarrow \infty} D(\omega^{*})=4 P_{\rm T} / \sigma_{v}^{2}$ holds is if $\omega^{*} \rightarrow 0$ and $\omega^{*} \theta \rightarrow \pi$, as $\theta \rightarrow \infty $.
\end{IEEEproof}
Theorem \ref{thm2} establishes that when $\theta$ is large we have an approximate closed-form solution for $\omega^{*} \approx \pi/\theta$ for any absolutely continuous sensing noise distribution.

\subsection{Finding the Optimum $\omega$ for Specific Noise Distributions} \label{sec: total_pow_cnst}
Theorem \ref{thm1} showed that $\omega^{*} \in (0, \pi/\theta)$ for a general class of distributions. Under more general conditions, Theorem \ref{thm2} establishes that $\omega^{*} \approx \pi/\theta$ when $\theta$ is large. To find $\omega^{*}$ exactly, we need to specify the sensing noise distribution through its CF, $\varphi_n(\omega)$. In what follows we describe how to find $\omega^{*}$ for several specific but widely used sensing noise distributions. We will assume throughout that the assumptions of Theorem \ref{thm1} ($\varphi_n^{'}(\omega)< 0$ for $\omega>0$) are satisfied so that $\omega^{*} \in (0, \pi / \theta)$, which holds for Gaussian, Cauchy and Laplacian distributions, among others. We will assume $\sigma_v^2>0$ throughout this subsection.
 
\subsubsection{Gaussian Sensing Noise}\label{sec: gauss_sens_noise}
In this case, we have $\varphi_n(\omega)=e^{- \omega^2 \sigma_{n}^{2} /2}$ so that $\varphi_n^2(\omega)=e^{- {\omega^2 \sigma_{n}^{2}}}$, where $\sigma_{n}^{2}$ is the variance of $n_i$. To simplify (\ref{eq:defl_coef_PT}) we substitute $\beta=\omega \theta$. Since $\omega \in (0, \pi / \theta)$ we have $\beta \in (0, \pi)$. Note that the value of $\omega$ that maximizes (\ref{eq:defl_coef_PT}) over $\omega$ is related to the $\beta$ that maximizes $D(\beta/ \theta)$ through the relation $\omega=\beta/ \theta$. Differentiating $D(\beta/ \theta)$ with respect to $\beta$, equating to 0 and simplifying we obtain,
\begin{equation}
\label{gauss_first_der}
G_{\rm G}(\beta) := \alpha - e^{- \frac{\sigma_n^2}{\theta^2} \beta^{2}} - \frac{ 2 \alpha \sigma_n^2}{\theta^2} \beta \tan \left (\frac {\beta}{2}\right)=0
\end{equation}
with $\alpha:= [1+ (\sigma_v^2 / P_{\rm T})]$. Equation \eqref{gauss_first_der} can not be solved in closed-form. However it does have a unique solution in $\beta \in (0, \pi)$ as shown in Appendix 1. The solution to \eqref{gauss_first_der}, $\beta^{*}_{\rm G}$, can be found numerically and the optimum $\omega$ for the Gaussian case is $\omega^{*}_{\rm G}= \beta^{*}_{\rm G} / \theta$.
\subsubsection{Cauchy Sensing Noise}\label{sec: cauchy_sens_noise}
In this case, $\varphi_n(\omega)=e^{- \gamma \omega}$ so that $\varphi_n^2(\omega)=e^{- 2 \gamma \omega}$ where $\gamma$ is the scale parameter of the Cauchy distribution. It is well known that no moments of this distribution exists. Substituting $\varphi_n(\omega)$ in $D(\omega)$ and letting $\beta=\omega \theta$ we have,
\begin{equation}
\label{cauchy_def_coeff}
D\left (\frac {\beta}{\theta}\right)  = \frac{[1 - \cos (\beta)]} {[\alpha e^{\frac{2 \gamma}{\theta} \beta} - 1] }
\end{equation}
with $\alpha:= [1+ (\sigma_v^2 / P_{\rm T})]$ and $\beta \in (0, \pi)$. It can be verified that the equation \eqref{cauchy_def_coeff} has a unique maximum over $\beta \in (0, \pi)$ as shown in Appendix 2. The $\beta^{*}_{\rm C}$ that maximizes \eqref{cauchy_def_coeff} can be found numerically and $\omega^{*}_{\rm C}= \beta^{*}_{\rm C} / \theta$.

When $\sigv/\powcnst $ is sufficiently large (i.e., the low channel SNR regime) compared to $[1-\varphi_n^{2} (\omega)]$ in $D(\omega)$, the problem in (\ref{eq:defl_coef_PT}) can be transformed into maximizing $\varphi_n^{2} (\omega)[1 - \cos (\omega \theta)]$ over $\omega \in (0, \pi/\theta)$. In this low channel SNR regime, we have a closed form solution for the Cauchy case:
\begin{equation}
\label{cauchy_low_ch_snr}
\omega^{*}_{\rm C}=\frac{2}{\theta} \tan^{-1}\frac{\theta}{2 \gamma}.
\end{equation}
If we let $\theta \rightarrow \infty$ in \eqref{cauchy_low_ch_snr}, we get $\omega^{*}_{\rm C}=\pi /\theta$ which agrees with Theorem \ref{thm2}.

\subsubsection{Laplace Sensing Noise}\label{sec: lapl_sens_noise}
In this case, we have $\varphi_n(\omega)=1/({1+b^2 \omega^2})$ and $b^2:=\sigma_n^2/2$. Substituting this in $D(\omega)$ and letting $\beta=\omega \theta$, and differentiating $D(\beta/ \theta)$ with respect to $\beta$, equating to 0 and simplifying we get,
\begin{equation}
\label{lapl_first_der}
G_{\rm L}(\beta) := \left [1 + \frac{b^2}{\theta^2} \beta^2 \right ]^2 - \frac{ 4 b^2}{\theta^2} \beta \left [1+ \frac{b^2}{\theta^2} \beta^2 \right ] \tan \left (\frac {\beta}{2}\right) - \left (\frac {1}{\alpha}\right) =0
\end{equation}
with $\alpha:= [1+ (\sigma_v^2 /P_{\rm T})]$. It can be easily verified that equation \eqref{lapl_first_der} has a unique solution in $\beta \in (0, \pi)$ as shown in Appendix 3. The $\beta^{*}_{\rm L}$ that solves \eqref{lapl_first_der} can be found numerically and $\omega^{*}_{\rm L}= \beta^{*}_{\rm L} / \theta$.

\subsubsection{Uniform Sensing Noise}\label{sec: uni_sens_noise}
For the uniform sensing noise, we have $\varphi_n(\omega)= \sin(\omega a) / {\omega a}$, where  $\sigma_n^2=a^2/3$. Substituting $\varphi_n(\omega)$ in \eqref{eq:defl_coef_PT} and letting $\beta=\omega a$ for convenience we have
\begin{equation}
\label{eq:def_coef_unif_noise} 
D(\beta)=\frac{ \left [1 - \cos \left( \frac{\beta \theta}{a} \right) \right] } { [\alpha \beta^2 \csc^2(\beta) -1]} = C(\beta) \left [1 - \cos \left( \frac{\beta \theta}{a} \right) \right]
\end{equation}
where $C(\beta):=1/[\alpha \beta^2 \csc^2(\beta) -1]$. Writing $D^{'}(\beta)=0$ we get
\begin{equation}
\label{eq:first_der_unif} 
\left [\alpha \beta^2 \csc^2(\beta) -1 \right ] - \alpha \beta \left [\frac{2 a }{\theta} \tan \left (\frac {\theta}{2a} \beta \right) \right ] \csc^2(\beta) [1-\beta \cot(\beta)]=0
\end{equation}
with $\alpha:= [1+ (\sigma_v^2 / P_{\rm T} )]$. Theorem \ref{thm1} does not apply for the uniform sensing noise. However if $\theta/a \geq 2$, then using $C(\beta) \geq C(\beta + k \pi)$, $k=1, 2, \ldots,$ and using the periodicity of $[1 - \cos (\beta \theta /a)]$, we can show that $\beta^{*}_{\rm U} \in (0, \pi a/\theta]$. Following similar arguments to the Laplacian noise case, it can be shown that there is only one stationary point in $(0, \pi a/\theta]$ which corresponds to the global maximum. The $\beta^{*}_{\rm U}$ that solves \eqref{eq:first_der_unif} can be found numerically and therefore, $\omega^{*}_{\rm U}=\beta^{*}_{\rm U}/a$. On the other hand if $\theta/a < 2$, multiple local maxima are possible in $\beta \in (0, \pi a/\theta]$ and \eqref{eq:first_der_unif} can have multiple solutions. In this case, that $\beta^{*}_{\rm U}$ which yields the largest value for $D(\beta)$ in \eqref{eq:def_coef_unif_noise} should be chosen.

\subsection{Per-sensor Power Constraint or high Channel SNR}\label{sec: per_sensor_pow_cnst}
We now consider the DC under the per-sensor power constraint. In this setting, as $L \rightarrow \infty$, $P_{\rm T} \rightarrow \infty$ which makes $ (\sigma_{v}^{2} / P_{\rm T}) \rightarrow 0$. Therefore the DC for the per-sensor constraint when $L$ is large becomes: 
\begin{eqnarray}
\label{eq:defl_coef_pspc}
D_{\rm pspc}(\omega)= \frac{2 \varphi_n^{2} (\omega)[1 - \cos (\omega \theta)]} {\left [1-\varphi_n^{2} (\omega) \right]}.
\end{eqnarray}
Equation \eqref{eq:defl_coef_pspc} can also be interpreted as the DC when $\sigma_v^2=0$ for any finite $L$. In what follows, we characterize $\omega^{*}$ in this per-sensor constraint regime, which effectively amounts to the removal of $(\sigma_v^2 / P_{\rm T} )$ from \eqref{eq:defl_coef_PT}. In this case there is not necessarily a $\omega^{*}$ that attains the maximum in \eqref{eq:omega_opt}. Our first result reveals that \eqref{eq:defl_coef_pspc} can be made large by choosing $\omega$ sufficiently close to zero when $n_i$ are Gaussian, and yields an interesting relationship between the DC and the Fisher information.
\begin{thm} \label{thm_opt_w_gauss}
When $n_i$ are Gaussian,
\begin{equation}
\label{eq:per_sensor_ineq1}
\sup_{\omega>0} D_{\rm pspc}(\omega)= \frac{\theta^2}{\sigma_{n}^{2}} = \lim_{\omega \rightarrow 0} D_{\rm pspc}(\omega)
\end{equation}
\end{thm}
\begin{IEEEproof}
We begin with the inequality $[1-\cos(\omega \theta)] \leq \omega^2 \theta^2 / 2$. Consider \cite[eqn (1)]{chinesephysics}, which using the fact that $\varphi_n(\omega)$ is real-valued, reveals $\varphi_n^2(\omega) \leq  (1+\varphi_n(2 \omega))/2$. Using these two inequalities we can write the following:
\begin{equation}
\label{eq:per_sensor_ineq1}
\frac{1}{D_{\rm pspc}(\omega)} \geq  \frac{[1 - \varphi_n(2 \omega)]}{2 \omega^2 \varphi_n^2(\omega)  \theta^2 } .
\end{equation}
Now from \cite[eqn (2)]{chinesephysics} with the fact that $\varphi_n(\omega)$ is real-valued, we have:
\begin{equation}
\label{eq:per_sensor_ineq2}
\frac{[1 - \varphi_n(2 \omega)]}{2 \omega^2 \varphi_n^2(\omega) } \geq \frac{1}{J}
\end{equation}
where $J$ is the Fisher information of $n_i$ with respect to a location parameter \cite[eqn (8)]{zamir} (i.e., the Fisher information in $x_i$ about $\theta$). Combining \eqref{eq:per_sensor_ineq1} and \eqref{eq:per_sensor_ineq2} we have:
\begin{equation}
\label{eq:per_sensor_ineq3}
\frac{D_{\rm pspc}(\omega)}{\theta^2} \leq J= \frac{1}{\sigma_{n}^{2}}
\end{equation}
where the equality follows from the fact that for Gaussian random variables the Fisher information is given by the inverse of the variance. Now, we also see that using l'H\^{o}spital's rule on \eqref{eq:defl_coef_pspc}, $\lim_{\omega \rightarrow 0} D_{\rm pspc}(\omega)=  \theta^2 / {\sigma_{n}^{2}}$, which shows that the inequality in \eqref{eq:per_sensor_ineq3} can be made arbitrarily tight establishing $\sup_{\omega > 0} D_{\rm pspc}(\omega)=  \theta^2 / {\sigma_{n}^{2}}$.
\end{IEEEproof}
The proof of Theorem \ref{thm_opt_w_gauss} also reveals an interesting inequality between the DC and the Fisher information, which of course is related to the Cram\'{e}r-Rao bound for unbiased estimators. So for the per-sensor power constraint case with Gaussian noise, $\omega$ should be chosen as small as possible for the best performance and it does not depend on the value of $\theta$. 

For the Laplacian case, the solution is similar to the Gaussian case. It can be easily verified that, with $(\sigma_{v}^{2} / P_{\rm T})=0$, $D^{'}_{\rm pspc}(\omega)<0$ over $\omega \in (0, \pi/\theta)$. This means that $D_{\rm pspc}(\omega)$ is monotonically decreasing with $\omega$ which implies that $\omega$ should be chosen arbitrarily small.

On the other hand, when $n_i$ are Cauchy distributed, then $\varphi_n(\omega)=e^{-\gamma \omega}$. Substituting in \eqref{eq:defl_coef_pspc} and using l'H\^{o}spital's rule we observe that  $\lim_{\omega \rightarrow 0} D_{\rm pspc}(\omega)=0$ for Cauchy sensing noise. This implies that, for the Cauchy sensing noise with per-sensor power constraint, smaller values of $\omega$ should be avoided for reliable detection to be possible.

\subsection{Analysis of the DC for Non-homogeneous Sensors}\label{sec: Non-homogeneous}
Consider now the case where $n_i$ are independent with non-identical distributions. This could occur if $n_i$ have the same type of distribution 
(e.g. Gaussian) with different variances. Letting $\varphi_{n_i}(\omega)={\rm E}[e^{j \omega n_{i}}]$, the DC in \eqref{eq:defl_coef_def} becomes
\begin{equation} 
\label{eq:defl_coef_PT_non_iid}
D_L(\omega)= \frac{2 [1 - \cos (\omega \theta)] \left( L^{-1} \displaystyle\sum_{i=1}^{L} \varphi_{n_i}(\omega)\right)^2} {\left [1- L^{-1} \displaystyle\sum_{i=1}^{L} \varphi_{n_i}^{2}(\omega) + \frac{\sigma_{v}^{2}}{P_{\rm T}}\right]}
\end{equation} 
which is now a function of $L$ unlike in \eqref{eq:defl_coef_PT}, and reduces to \eqref{eq:defl_coef_PT} if $\varphi_{n_i}(\omega)=\varphi_n(\omega)$, as in the iid case. We now study the conditions on the variances $\sigma_{i}^{2}:={\rm var}(n_i)$ for $\lim_{L \rightarrow \infty} D_L(\omega)=0$ for all $\omega>0$. When this asymptotic DC is zero for all $\omega>0$, the interpretation is that there is no suitable choice for $\omega>0$. The following result establishes that if the sensing noise variances are going to infinity, the asymptotic DC is zero for all $\omega>0$, indicating a regime where reliable detection is not possible.

\begin{thm} \label{thm_non_iid}
Let  $\varphi_{n_i}(\omega)= \varphi_n(\sigma_{i} \omega )$ for some CF $\varphi_n(\omega)$ where $n$ has an absolutely continuous distribution. Suppose also that $\lim_{i \rightarrow \infty} \sigma_{i}=\infty$. Then $\lim_{L \rightarrow \infty} D_L(\omega)=0$ for all $\omega>0$.
\end{thm}
\begin{IEEEproof}
Clearly the denominator of \eqref{eq:defl_coef_PT_non_iid} is bounded between $(\sigma_{v}^{2} / P_{\rm T})$ and $(1+\sigma_{v}^{2} / P_{\rm T})$. Therefore, it suffices to show that $L^{-1} \sum_{i=1}^{L} \varphi_{n_i}(\omega) = L^{-1} \sum_{i=1}^{L} \varphi_n( \sigma_{i} \omega ) \rightarrow 0$ as $L \rightarrow \infty$. Since $n$ has an absolutely continuous distribution, $\lim_{x \rightarrow \infty} \varphi_n(x)=0$, and because $\lim_{i \rightarrow \infty} \sigma_i = \infty$, it follows that $\lim_{i \rightarrow \infty} \varphi_n(\sigma_i \omega) =0$ for $\w>0$. From \cite[pp. 411]{porat1994} we know that if a sequence satisfies $\lim_{i \rightarrow \infty} a_i = 0$ then its partial sums also satisfy $\lim_{L \rightarrow \infty} L^{-1}\sum_{i=1}^L a_i=0$, which gives us the proof when applied to the sequence $\varphi_n(\sigma_i \omega)$.
\end{IEEEproof}
If, instead of $\sigma_{i}^{2} \rightarrow \infty$ as $i \rightarrow \infty$, the variances $\sigma_{i}^{2}$ are bounded, we can show the existence of an $\omega>0$ for which $\lim_{L \rightarrow \infty} D_L(\omega)>0$ which is done next. 
\begin{thm} \label{varbounded}
Let ${\rm var}(n_i)$ exist for all $i$ and $\sigma_{\rm max} := \sup_{i} ({\rm var}(n_i))^{1/2}$ be finite. Then any $0 < \w < \sqrt{2}/\sigma_{\rm max}$ satisfies $\lim_{L \rightarrow \infty} D_L(\omega)>0$.
\end{thm}
\begin{IEEEproof}
To show $\lim_{L \rightarrow \infty} D_L(\omega)>0$ for $\omega>0$, it suffices to show that $L^{-1} \sum_{i=1}^{L} \varphi_{n_i}(\omega)>0$ for $\omega>0$. From \cite[pp. 89]{ushakov1999} we have $\varphi_{n_i}(\w) \geq 1- \sigma_{i}^2 \w^2/2$ for any CF with finite variance. Therefore, $L^{-1} \sum_{i=1}^{L} \varphi_{n_i}(\omega) \geq  1- (L^{-1}\sum_{i=1}^L \sigma_{i}^2) \w^2/2 \geq 1- \sigma_{\rm max}^2 \w^2/2 >0$ where the last inequality holds provided that $\w < \sqrt{2}/\sigma_{\rm max}$.
\end{IEEEproof}
This shows that if the noise variances are bounded, there exists (a small enough) $\omega$ that yields a strictly positive asymptotic DC, establishing that there is a choice of $\omega$ that enables reliable detection.

\section{Fading Channels} \label{sec:fading_channels}
Suppose that the channel connecting the $i^{th}$ sensor and the FC has a fading coefficient $h_i:= |h_i| e^{j\phi_i}$ normalized to satisfy  E$[|h_i|^2]=1$. If the sensors do not know or utilize their local channel information, and the fading has zero-mean (E$[h_i]=0$), then the performance over fading channels is poor because the DC in \eqref{eq:defl_coef_def} becomes zero due to law of large numbers and reliable detection is not possible. On the other hand, if the $i^{th}$ sensor corrects for the channel phase before transmission, using local channel phase information, the received signal under the TPC becomes
\begin{equation}
\label{eq:recd_signal_fading}
y_L = e^{j \omega \theta} \sqrt{\frac{P_{\rm T}}{L}}\displaystyle\sum_{i=1}^{L} |h_i| e^{j \omega n_{i}} + v \;,
\end{equation}
where we focus on the iid sensing noise case to highlight the effect of fading even though the non-homogeneous case can also be easily pursued. The phase correction does not change the constant power nature of the transmission. By calculating the expectations in \eqref{eq:defl_coef_def}, for the signal model in (\ref{eq:recd_signal_fading}), the DC in the presence of fading is given by: 
\begin{equation}
\label{eq:defl_coef_fading}
D(\omega)= \frac{2 ({\text{E}}[|h_i|])^{2} \varphi_n^{2} (\omega)[1 - \cos (\omega \theta)]} {\left [1 - ({\text{E}}[|h_i|])^{2} \varphi_n^{2} (\omega) + \frac{\sigma_{v}^{2}}{P_{\rm T}}\right]} .
\end{equation}
We see that in case of fading, the term $\varphi_n^{2} (\omega)$ is scaled by the factor $({\text{E}}[|h_i|])^{2}$ in the DC expression. Since ${\text{E}}[|h_i|^2]=1$, using Jensen's inequality, the factor $({\text{E}}[|h_i|])^{2} <1$ unless $|h_i|$ is deterministic in which case it is one. Comparing \eqref{eq:defl_coef_PT} and \eqref{eq:defl_coef_fading} we have, with $({\text{E}}[|h_i|])^{2} <1$, the numerator of \eqref{eq:defl_coef_fading} is decreased and the denominator of \eqref{eq:defl_coef_fading} is increased, leading to a reduction in DC and thus fading has a detrimental effect on the detection performance, as expected.

Note that if the optimization of the DC is desired in the fading case, the factor $({\text{E}}[|h_i|])^{2}$ in the denomenator of \eqref{eq:defl_coef_fading} affects the optimum $\omega$ value. Theorem \ref{thm1} can be proved for the fading case as well with $C(\omega):=2 ({\text{E}}[|h_i|])^{2} \varphi_n^{2} (\omega) / [1- ({\text{E}}[|h_i|])^{2} \varphi_n^{2} (\omega) + \sigma_{v}^{2} / P_{\rm T} ]$ which is still decreasing with $\omega$ if $\varphi_n(\omega)$ is. Therefore the conclusion of Theorem \ref{thm1}, namely, $\omega^{*} \in (0, \pi/\theta)$, does not change. The procedure to find the $\omega^{*}$ under the TPC for Gaussian, Cauchy and Laplacian is the same as described in Sections \ref{sec: gauss_sens_noise}, \ref{sec: cauchy_sens_noise} and \ref{sec: lapl_sens_noise} respectively. The equations \eqref{gauss_first_der}, \eqref{gauss_sec_der}, \eqref{gauss_sec_der2} and \eqref{cauchy_first_der1} remain valid with the exponentials in these equations scaled by the factor $({\text{E}}[|h_i|])^{2}$. The equations \eqref{lapl_first_der}, \eqref{lapl_sec_der} and \eqref{lapl_sec_der2} for the Laplacian case also remain valid except the term $1/\alpha$ in \eqref{lapl_first_der} scaled by 
$({\text{E}}[|h_i|])^{2}$.

We note that if sensors have imperfect knowledge of the phase, $|h_i|$ will be replaced by $|h_i| e^{j \tphii}$ where $\tphii$ is the phase error. Clearly this error can also be subsumed in (\ref{eq:recd_signal_fading}) as replacing $\omega n_i$ with $\omega n_i + \tphii$ which changes the sensing noise by a term independent of $\omega$. This establishes the interesting fact that phase error over fading channels can be treated as a change in sensing noise distribution.

\section{Asymptotic Performance and Optimization of $\omega$ based on error exponent} \label{sec:error_exponent}
The error exponent in a distributed detection system is a measure of how fast the probability of error goes to zero as $L \rightarrow \infty$. Mathematically error exponent is defined as:
\begin{equation}
\label{eq:err_exp}
-\lim_{L \rightarrow \infty} \frac{ \log P_{\rm e}(\omega) } {L} .
\end{equation}
Large deviation theory \cite{Bucklew,Hollander} provides a systematic procedure to calculate the error exponent which is briefly reviewed next. Let $Y_L$ be a sequence of random variables without any assumptions on their dependency structure and let $M(t)= \lim_{L \rightarrow \infty} (1/L) \log {\rm E} \lbrace e^{t Y_L}\rbrace$ exist and is finite for all $t \in R$. Define
\begin{equation}
\label{eq:gartner_ellis}
\varepsilon(z) = - \lim_{L \rightarrow \infty} \frac{1}{L}\log \Pr \left[ Y_L < z \right] \;,
\end{equation}
where $z$ is the threshold and $Y_L$ is the test statistic of a detector. G$\ddot{\rm a}$rtner-Ellis Theorem \cite[pp. 14]{Bucklew} states that $\varepsilon(z)$ in \eqref{eq:gartner_ellis} can be calculated using,
\begin{equation}
\label{eq:gartner_ellis2}
\varepsilon(z) = \sup_{t \in R} [ t z - M(t)] \;,
\end{equation}
where 
\begin{equation}
\label{eq:mgf}
M(t)=\lim_{L \rightarrow \infty} \frac{1}{L}\log {\rm E} \lbrace e^{t Y_L}\rbrace  .
\end{equation}
We will now use the G$\ddot{\rm a}$rtner-Ellis Theorem with $Y_L$ replaced by $Z_L(\omega)$ in \eqref{eq:prob_error_bayesian2} and $z=0$. Letting $M_{\omega}(t):= \lim_{L \rightarrow \infty} (1/L) \log {\rm E} \lbrace e^{t Z_L(\omega)}\rbrace$, and $\varepsilon_{\omega}(z) = \sup_{t \in R} [ t z - M_{\omega}(t)]$ we have the following theorem which relates the error exponent to the CF $\varphi_n(\omega)$ of the sensing noise distribution.
\begin{thm} \label{thm3}
For the detector in (\ref{eq:detector_gaussian}), the error exponent in \eqref{eq:err_exp} is $\varepsilon_{\omega}(0) = - \inf_{t \in R} M_{\omega}(t)$ where
\begin{align}
\label{eq:err_m_t} 
M_{\omega}(t)=\log  \left [ \displaystyle  I_{0} \left (2 \sin \left (\frac{\omega \theta}{2} \right) t \right)+ 2 \sum_{k=1}^{\infty} I_{k} \left (2 \sin \left (\frac{\omega \theta}{2} \right) t \right)  \varphi_n(k \omega) \cos \left( k \left (\frac{\pi}{2}  - \frac{\omega \theta}{2} \right)\right ) \right ]  \nonumber \\ + \left [ \frac{ t^{2}  \sigma_{v}^{2}(1-\cos (\omega\theta))}{2P_{\rm T}} \right ]
\end{align}
where $I_{k}(t)$ is the modified Bessel function of the first kind.
\end{thm}
\begin{IEEEproof}
Please see Appendix 4.
\end{IEEEproof} 
It is well known that the function $M_{\omega}(t)$ is convex in $t$ \cite{Bucklew}. Therefore the supremum in (\ref{eq:gartner_ellis2}) can be found efficiently for $z=0$. The $t^{*}$ that maximizes (\ref{eq:gartner_ellis2}) satisfies $M_{\omega}^{'}(t^{*})=0$ which can be found by convex methods with geometric convergence \cite{Boyd}.

In addition to the error exponent, it is also possible to approximate $P_{\rm e}(\omega)$ using the function $\varepsilon_{\omega}(z)$. In fact Bahadur and Rao  \cite[pp. 10]{Hollander} have proved that this probability can be approximated using the error exponent and is given by:
\begin{equation}
\label{eq:ber_err_exp}
P_{\rm e}(\omega)= \frac{1}{ \sqrt{2 \pi {\hat{\sigma}}^2_{\omega} } } e^{- L \varepsilon_{\omega}(0) \left( 1 + o(1)\right)} \;,
\end{equation}
as $L \rightarrow \infty$ and ${\hat{\sigma}}^2_{\omega}:= [\varepsilon_{\omega}^{'}(0)]^2 / [\varepsilon_{\omega}^{''}(0)]$. The quantities $\varepsilon_{\omega}^{'}(0)$ and $\varepsilon_{\omega}^{''}(0)$ are the first and second derivatives of $\varepsilon_{\omega}(z)$ at $z=0$ respectively, and can be calculated from the following equations \cite[pp. 121]{Boyd}:
\begin{equation}
\label{eq:eta_first_der}
\varepsilon_{\omega}^{'}(0) = t^{*} \;,
\end{equation}
\begin{equation}
\label{eq:eta_2nd_der}
\varepsilon_{\omega}^{''}(0) = \frac{1}{M_{\omega}^{''}(t^{*})} . 
\end{equation}

The error exponent given in Theorem \ref{thm3} is a function of $\omega$ and let us denote it by $\varepsilon_{\omega}$ for convenience. It will be illustrated in Section \ref{sec: simulations} that the values of $\omega$ that minimizes $P_{\rm e}(\omega)$ is closely predicted by the value obtained by maximizing $D(\omega)$ or $\varepsilon_{\omega}$. We will also examine in the simulations in Section \ref{sec: simulations} how
accurately \eqref{eq:ber_err_exp} can be used to approximate $P_{\rm e}(\omega)$ for finite $L$.

\section{Non-Gaussian Channel Noise} \label{sec: nonGaussianChannel}
We have so far assumed that the channel noise as Gaussian. However, we verified that the detector in \eqref{eq:detector_gaussian} works well even if the channel noise is mixed Gaussian, uniform or Laplacian. The channel noise distribution will only affect the error exponent through the second term in \eqref{eq:err_m_t}. Using this, the effect of different channel noise distributions we considered are briefly sketched below.

We considered the case of mixed Gaussian having two different variances drawn from a Bernoulli distribution. Let $p_0$ be the probability that the samples drawn from the mixture have variance $\sigma_{v_{0}}^2$ and $p_1=1-p_0$ be the probability corresponding to $\sigma_{v_{1}}^2$ and let $\sigma_{v_{1}}^2 > \sigma_{v_{0}}^2$. In this case, we found that the error exponent is affected only by the larger variance in the mixture. While using G$\ddot{\rm a}$rtner-Ellis Theorem to calculate $M_{\omega}(t)$, the second term in \eqref{eq:err_m_t} for the mixed Gaussian becomes $\lim\limits_{L \rightarrow \infty} L^{-1}\log  \left [ p_0 \exp \left( { t^{2}  \sigma_{v_{0}}^{2}(1-\cos (\omega\theta))} / {2P_{\rm T}} \right)  +  p_1 \exp \left( { t^{2}  \sigma_{v_{1}}^{2}(1-\cos (\omega\theta))}/{2P_{\rm T}} \right)  \right]$ and this limit evaluates to $ [{ t^{2}  \sigma_{v_{1}}^{2}(1-\cos (\omega\theta))} / {2P_{\rm T}}]$ which proves that only the larger variance $\sigma_{v_{1}}^{2}$ in the mixture affects the error exponent.

For the uniform channel case, interestingly we found that the second term in \eqref{eq:err_m_t} evaluates to 0 and thus proving that the error exponent is not impacted by the uniform channel noise. We do not include the straightforward derivation due to lack of space. We will discuss the performance of the mixed Gaussian and Laplacian cases in Section \ref{sec: simnonGaussianChannel}.

\section{Simulations} \label{sec: simulations}
We define the sensing and channel SNRs as $\rho_s:= \theta^2 / \sigma_{n}^{2}$, $\rho_c:={P_{\rm T}}/{\sigma_{v}^{2}}$ and assume $P_1=P_0=0.5$ throughout. Note also that $\rho=P_{\rm T}/L$ is the power at each sensor as defined in Section \ref{sec: model}.

\subsection{Effect of $\omega$ on Performance} \label{subsec: pe_w_eff}
\begin{figure}[tb]
\begin{minipage}{1\textwidth}
\centering
\begin{center}
\includegraphics[height=9cm,width=12cm]{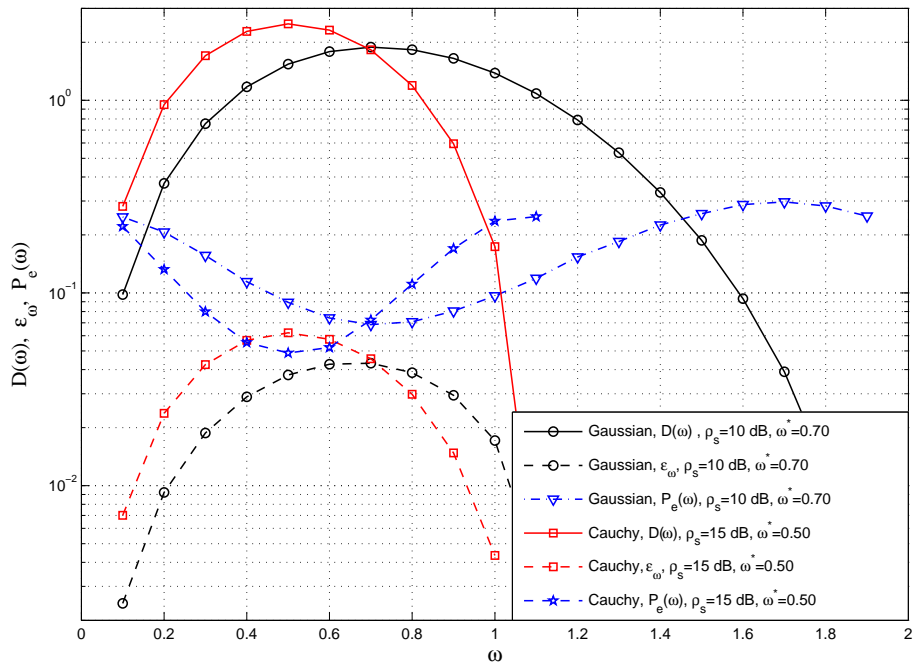}
\caption{Total Power Constraint, $D(\omega), \varepsilon_{\omega}, P_{\rm e}(\omega)$ vs $\omega$: $\rho_s$=10, 15 dB, $\rho_c$=-10 dB,  $L$=20}\label{fig: Pe_w_D_Gaussian_Cauchy}
\end{center}
\end{minipage}
\end{figure}

We begin by comparing the optimized $\omega$ values using $D(\omega)$, $\varepsilon_{\omega}$ and $P_{\rm e}(\omega)$ for the TPC. The values of $\omega^{*}>0$ obtained by maximizing the error exponent $\varepsilon_{\omega}$ and the DC $D(\omega)$ were found to be very close over the entire range of $P_{\rm T}$. Figure \ref{fig: Pe_w_D_Gaussian_Cauchy} shows the plots of $D(\omega), \varepsilon_{\omega}$, and $P_{\rm e}(\omega)$ vs $\omega$ for Gaussian and Cauchy sensing noise distributions where the $P_{\rm e}(\omega)$ plot is obtained using Monte-Carlo simulations. The different $\omega^{*}$ values in Figure \ref{fig: Pe_w_D_Gaussian_Cauchy} correspond to the best $\omega$ values obtained by optimizing $D(\omega)$, $\varepsilon_{\omega}$ and $P_{\rm e}(\omega)$ respectively. It is interesting to see that the $\omega^{*}$ that minimizes $P_{\rm e}(\omega)$ is very close to that which maximizes $D(\omega)$ and $\varepsilon_{\omega}$. For Laplacian and Uniform sensing noises (not shown), the same trends were observed. 

\begin{figure}[tb]
\begin{minipage}{1\textwidth}
\centering
\begin{center}
\includegraphics[height=9cm,width=12cm]{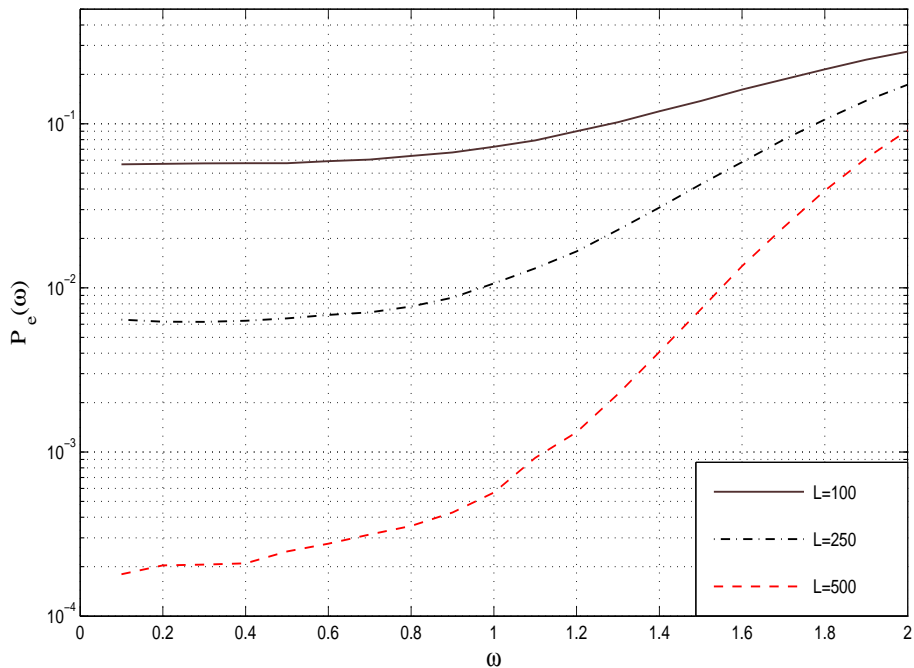}
\caption{Per-sensor Power Constraint, Gaussian, $P_{\rm e}(\omega)$ vs $\omega$: $\rho_s$=-10 dB, $\rho$=10 dB}\label{fig: Pe_w_Gaussian_pspc}
\end{center}
\end{minipage}
\end{figure}

Figure \ref{fig: Pe_w_Gaussian_pspc} shows the performance under per-sensor power constraint with large $L$. It is observed that smaller $\omega$ yields better error probability. This agrees with our findings in Section \ref{sec: per_sensor_pow_cnst} where it was shown that $D_{\rm pspc}(\omega)$ can be made larger by choosing $\omega>0$ arbitrarily small. Since both Figures \ref{fig: Pe_w_D_Gaussian_Cauchy} and \ref{fig: Pe_w_Gaussian_pspc} verify that the choice of $\omega$ based on minimizing $P_{\rm e}(\omega)$ can be closely approximated by that which maximizes $D(\omega)$, in all subsequent simulations, we have used the $\omega^{*}$ values obtained by maximizing $D(\omega)$.

\subsection{Comparison against MAF and MDF Schemes} \label{subsec: comp_maf_mdf_sch}
\begin{figure}[tb]
\begin{minipage}{1\textwidth}
\centering
\begin{center}
\includegraphics[height=9cm,width=12cm]{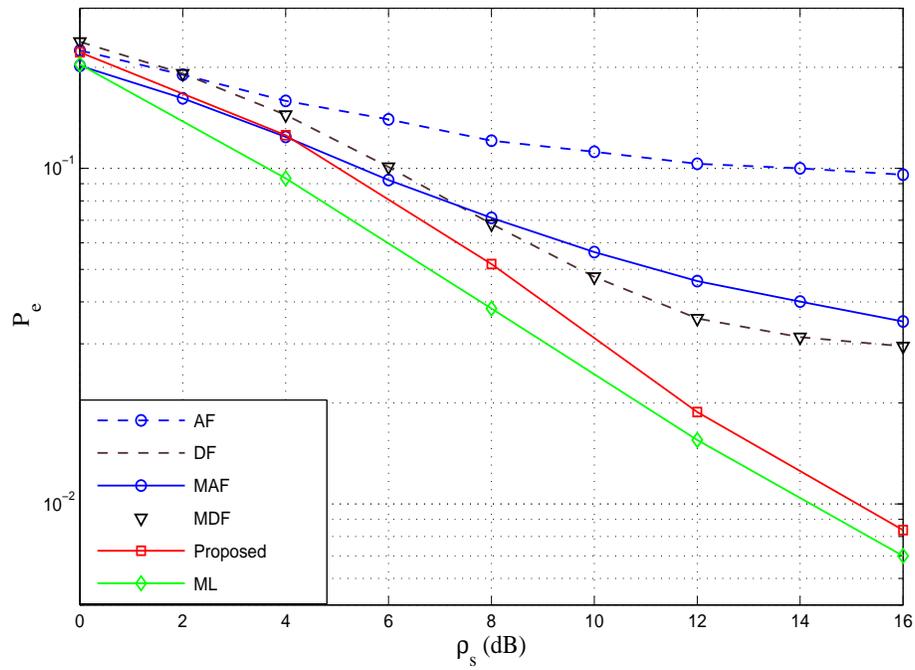}
\caption{Total Power Constraint, $P_{\rm e}$ vs $\rho_s$: $\rho$=-30 dB, $L$=60}\label{fig: MAF_MDF_PM2}
\end{center}
\end{minipage}
\end{figure}

\begin{figure}[tb]
\begin{minipage}{1\textwidth}
\centering
\begin{center}
\includegraphics[height=9cm,width=12cm]{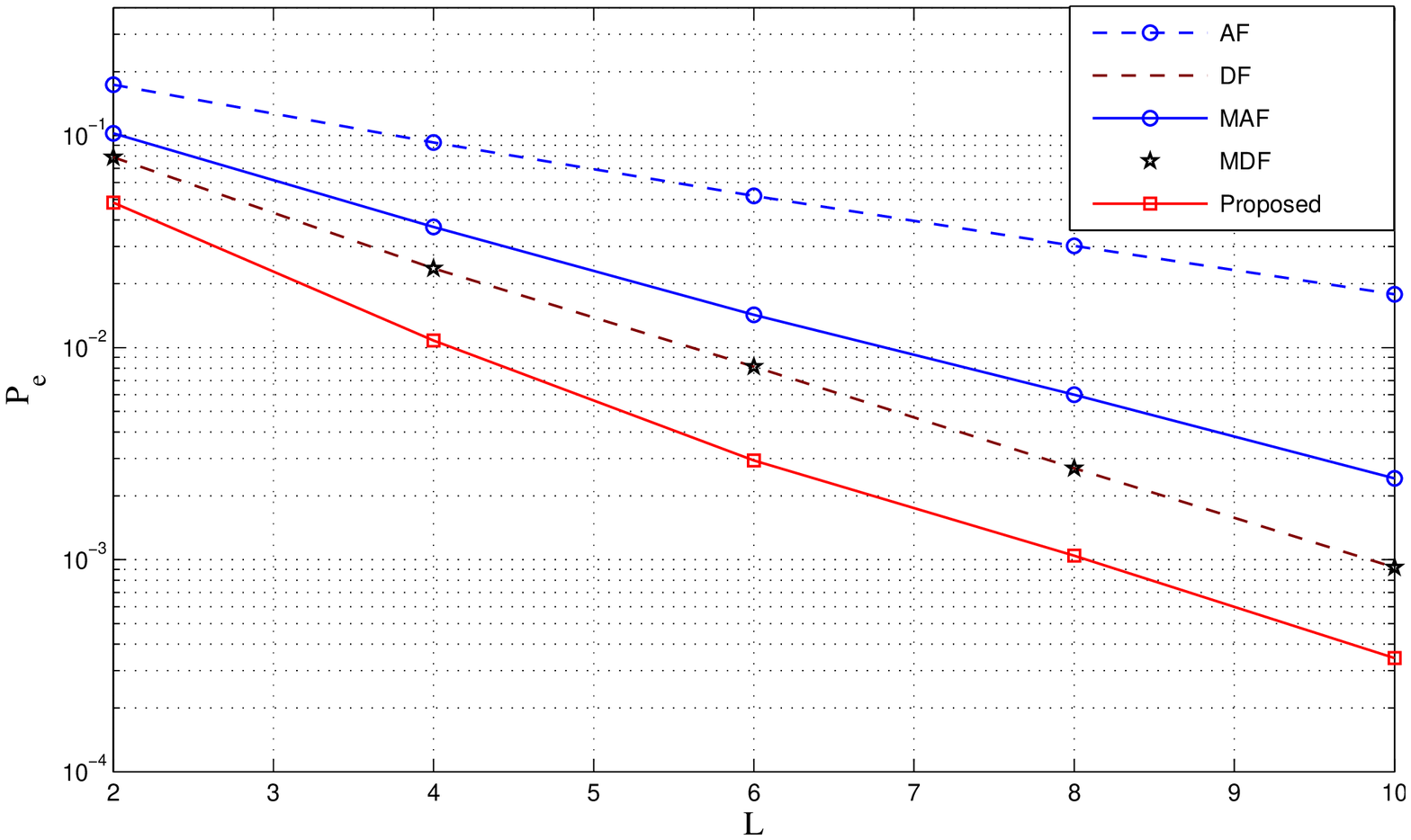}
\caption{Total Power Constraint, $P_{\rm e}$ vs $L$: $\rho_s$=15 dB, $\rho_c$=0 dB}\label{fig: Pe_comp_PM_others_L}
\end{center}
\end{minipage}
\end{figure}

In Figure \ref{fig: MAF_MDF_PM2}, the proposed scheme is compared under the TPC with the MAF and MDF schemes which have been shown in \cite{Evans_DD_MA} to outperform conventional amplify-and-forward (AF) and detect-and-forward (DF) schemes. We observe that the proposed scheme outperforms MAF when $\rho_s >$ 4 dB, and MDF for the entire range of $\rho_s$. The same trend was observed when $L$ is increased to 90 with an improvement in the detection error probability. The ML performance shown was obtained by the Monte-Carlo implementation of the ML detector and is computationally complex, but serves as a performance benchmark. Figure \ref{fig: Pe_comp_PM_others_L} shows the $P_{\rm e}$ performance versus $L$ under the TPC. Clearly the proposed scheme outperforms the AF, DF, MAF and MDF schemes consistently since $\rho_s =$ 15 dB. 

The proposed scheme requires the fine tuning of the transmission phase parameter $\omega$ either through optimizing the deflection coefficient or the error exponent. However, it should be noted that similar type of fine tuning is also required in the competing schemes such as the MAF or the MDF. We note that the proposed scheme is inferior to MAF at low sensing SNRs ($\rho_s <$ 4 dB). On the whole, the benefits of constant modulus signaling and improved performance at higher sensing SNRs make the proposed approach a viable alternative.

\subsection{Total Power Constraint: Different Noise Distributions} \label{subsec: tpc_diff_noises_comp}
\begin{figure}[tb]
\begin{minipage}{1\textwidth}
\centering
\begin{center}
\includegraphics[height=9cm,width=12cm]{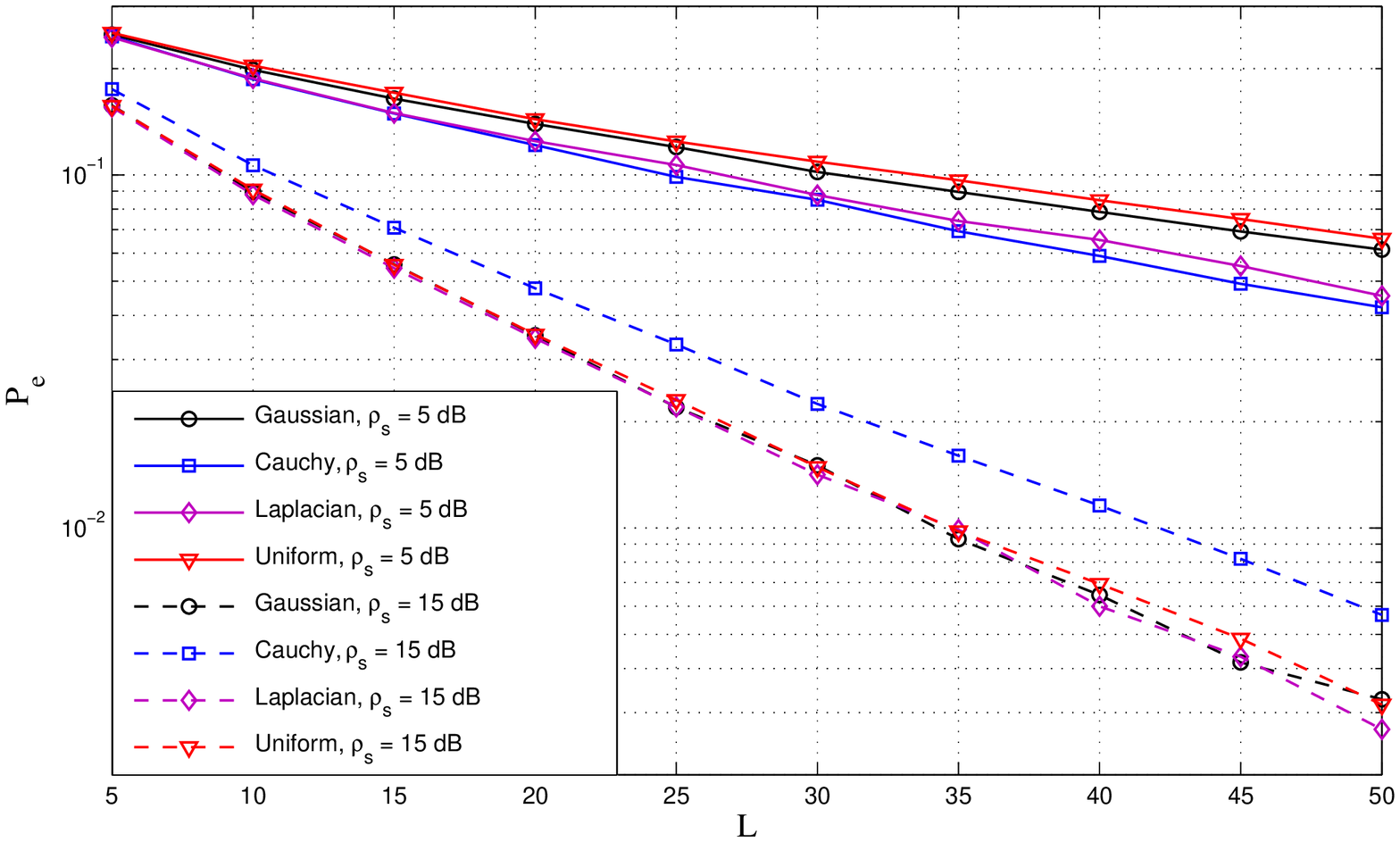}
\caption{Total Power Constraint, $P_{\rm e} $ vs $L$: $\rho_s$=5, 15 dB, $\rho_c$=-10 dB}\label{fig: pe_tsc_diff_dist_L}
\end{center}
\end{minipage}
\end{figure}

For the Total Power Constraint, Figure \ref{fig: pe_tsc_diff_dist_L} shows that Cauchy sensing noise results in better performance when $\rho_s$ is  low, and worse when $\rho_s$ is  high compared with other sensing noise distributions. This agrees with the fact that $D(\omega^{*})$ is smaller for Cauchy sensing noise when $\rho_s$ is high than other distributions and vice versa when $\rho_s$ is low. When $\rho_s$ is moderately high, we observe that Gaussian, Laplacian and Uniform distributions have identical performance if $\rho_c$ is very low for a wide range of $L$ as illustrated in the Figure \ref{fig: pe_tsc_diff_dist_L}. We found numerically that the similarity of the $P_{\rm e}(\omega)$ curves under different sensing noise distributions was also reflected in the corresponding $D(\omega)$ values where they were also verified to be similar. 

\begin{figure}[tb]
\begin{minipage}{1\textwidth}
\centering
\begin{center}
\includegraphics[height=9cm,width=12cm]{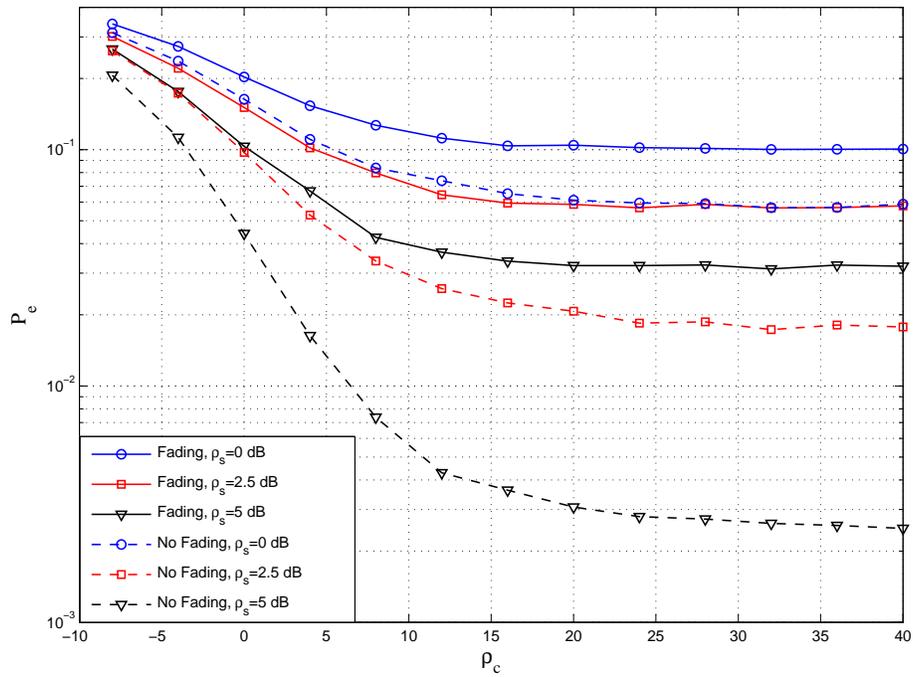}
\caption{Rayleigh flat fading, $P_{\rm e} $ vs $\rho_c$, $n_i$ Gaussian, $L$=10}\label{fig: Pe_TPC_Fading}
\end{center}
\end{minipage}
\end{figure}

Figure \ref{fig: Pe_TPC_Fading} compares the performance of the proposed scheme in the presence of Rayleigh flat fading between the sensors and the FC against without fading with the Gaussian sensing noise. Clearly, fading has a detrimental effect on the detection performance as argued in Section \ref{sec:fading_channels}. It is also observed that, in the presence of fading, $P_{\rm e}$ is not as sensitive to the increase in $\rho_s$ as that of the no fading case.

\subsection{Error Exponent} \label{subsec: err_exponent_comp_diff}
\begin{figure}[tb]
\begin{minipage}{1\textwidth}
\centering
\begin{center}
\includegraphics[height=9cm,width=12cm]{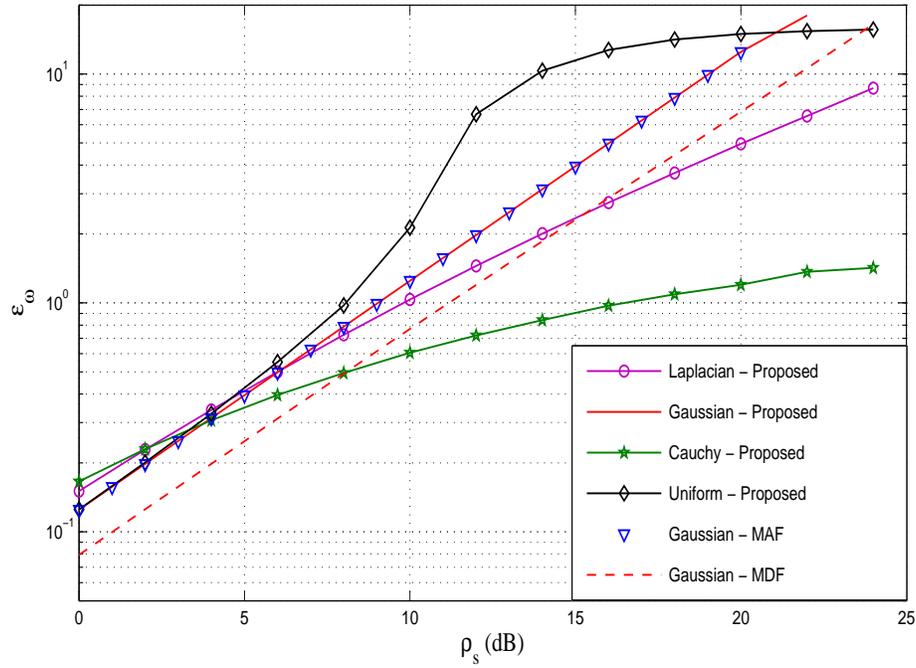}
\caption{Per-sensor Power Constraint, $\varepsilon_{\omega} $ vs $ \rho_s$}\label{fig: eta_psc_diff_dist}
\end{center}
\end{minipage}
\end{figure}

Figure \ref{fig: eta_psc_diff_dist} depicts the error exponent of the proposed scheme under the PSPC and illustrates its improvement with increase in $\rho_s$ for all the sensing noise distributions. Recall that $\sigma_{v}^{2}$ has no effect on the error exponent for the PSPC case since $(\sigma_{v}^{2} / P_{\rm T}) \rightarrow 0$ in \eqref{eq:err_m_t}. It is interesting to see that Cauchy sensing noise has a better error exponent than Gaussian, Laplacian and Uniform sensing noise distributions when $\rho_s \leq $ 4 dB while it is worse when $\rho_s>$ 4 dB. The error exponent with Gaussian sensing noise is better than that of Laplacian noise when when $\rho_s>$ 7.5 dB and the uniform distribution has a better error exponent than other sensing noise distributions when $\rho_s>$ 4 dB. The error exponent of the proposed scheme is compared with the error exponents of MAF and MDF schemes which were only derived for the Gaussian case (please see equations (24) and (25) in \cite{Evans_DD_MA}). It is seen that, for the PSPC case, the MAF scheme (whose error exponent is $\rho_s/8$) and the proposed scheme with optimum $\omega$ have identical error exponents leading us to conjecture that $\sup_{\omega} [- \inf_{t \in R} M_{\omega}(t)]=\rho_s/8$ when $n_i$ are Gaussian. The MDF error exponent is inferior compared to MAF and the proposed scheme.

\begin{figure}[tb]
\begin{minipage}{1\textwidth}
\centering
\begin{center}
\includegraphics[height=9cm,width=12cm]{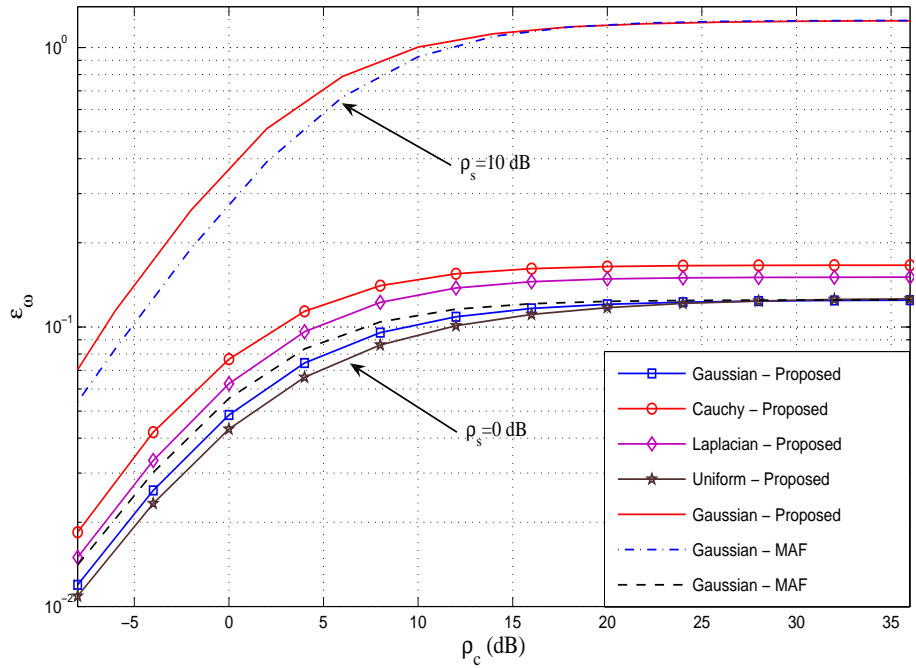}
\caption{Total Power Constraint, $\varepsilon_{\omega}$ vs $\rho_c$: $\rho_s$=0, 10 dB, $\sigma_{v}^{2}$=1}\label{fig: eta_tsc_diff_dist}
\end{center}
\end{minipage}
\end{figure}

Figure \ref{fig: eta_tsc_diff_dist} shows the error exponent under the TPC with $\rho_s=0$ dB. In this scenario, Cauchy sensing noise has the best error exponent since $\rho_s$ is low. This concurs with the fact illustrated in Section \ref{subsec: tpc_diff_noises_comp} that the DC of Cauchy is better at lower values of $\rho_s$ than other distributions and this was justified by the simulation results as shown in Figure \ref{fig: pe_tsc_diff_dist_L}. We found that when $\rho_s$ is increased, Cauchy becomes inferior to other noise distributions. For all the distributions, increasing $\rho_c$ results in an increase in the error exponent which becomes a constant beyond $\rho_c=$ 15 dB. This is because, for a given $\rho_s$, increasing $\rho_c$ combats the effect of channel noise, thereby improving the error exponent. However, the effect of sensing noise can not be overcome by increasing $\rho_c$ indefinitely. This can be seen from \eqref{eq:err_m_t} as well where the second term vanishes while the first term remains even for large $P_{\rm T}$. For the Gaussian case, we derived the error exponent of the MAF scheme under the TPC as $\varepsilon_{\rm MAF}= \theta^2 / 8 [\sigma_{n}^{2} +( \sigma_{v}^{2} (\sigma_{n}^{2} + P_0 P_1 \theta^2 )/P_{\rm T})]$. If $P_{\rm T} \rightarrow \infty$, this reduces to $\rho_s/8$ for the PSPC case. It is seen that with $\rho_s=$ 0 dB, the MAF scheme is better than the proposed method when $\rho_c < $ 15 dB. However, under the TPC, the error exponent of the proposed scheme was found to beat the MAF scheme when $\rho_s > $ 4.5 dB and an example plot is shown in Figure \ref{fig: eta_tsc_diff_dist} for $\rho_s = $ 10 dB. This crossover between the MAF and the proposed schemes is also reflected in their respective $P_{\rm e}$ performance curves approximately around the same $\rho_s$ value (please see Figure \ref{fig: MAF_MDF_PM2}). However, if $\rho_c$ is increased beyond 15 dB, we see that the error exponents of both the schemes become very close.

\subsection{Approximations of $P_{\rm e}(\omega)$ through $\varepsilon_{\omega}(z)$} \label{subsec: ber_predict}
\begin{figure}[tb]
\begin{minipage}{1\textwidth}
\centering
\begin{center}
\includegraphics[height=9cm,width=12cm]{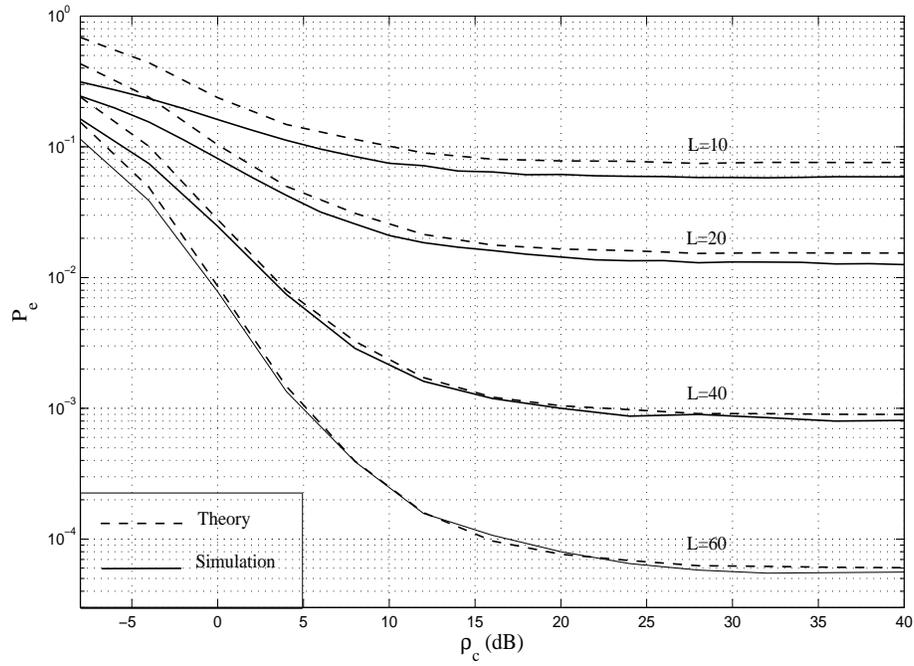}
\caption{Gaussian Sensing Noise: $P_{\rm e} $ vs $\rho_c$: $\rho_s$=0 dB, $L$=10, 20, 40, 60}\label{fig: gauss_tsc_ee_sim}
\end{center}
\end{minipage}
\end{figure}

\begin{figure}[tb]
\begin{minipage}{1\textwidth}
\centering
\begin{center}
\includegraphics[height=9cm,width=12cm]{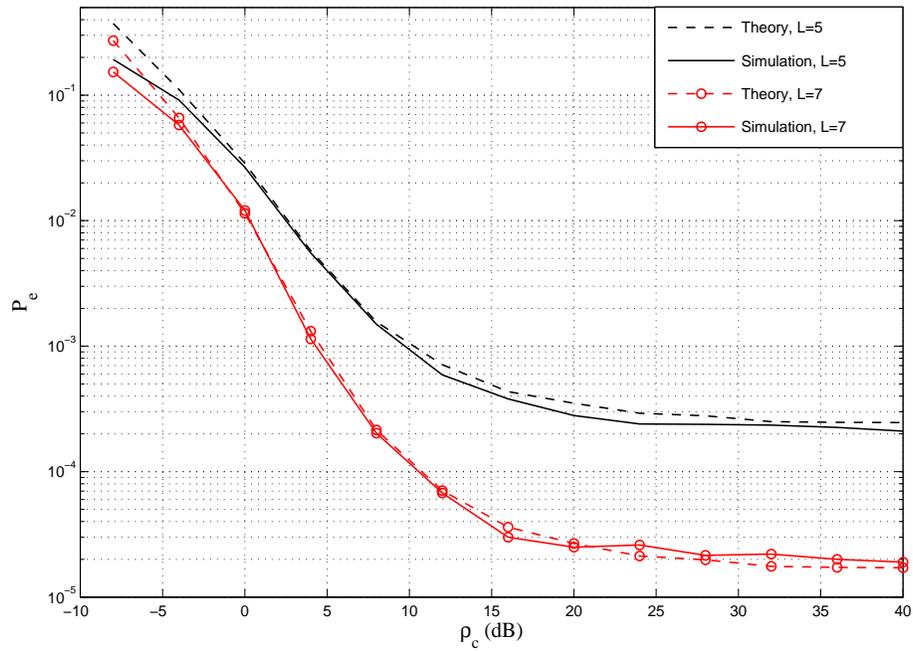}
\caption{Gaussian Sensing Noise: $P_{\rm e} $ vs $\rho_c$: $\rho_s$=10 dB, $L$=5, 7}\label{fig: gauss_tsc_ee_sim2}
\end{center}
\end{minipage}
\end{figure}

Equation \eqref{eq:ber_err_exp} provides an approximation of $P_{\rm e}(\omega)$ based on the error exponent. The expression in \eqref{eq:ber_err_exp} is found to match with the simulations when $\rho_c  > $ 0 dB and $\rho_s > $ -5 dB. Figures \ref{fig: gauss_tsc_ee_sim} and \ref{fig: gauss_tsc_ee_sim2} elucidate this behavior for Gaussian sensing noise distribution. Similar trends were observed for the other sensing noise distributions as well but are not shown due to space constraints. When $L$ is small, the gap between theory and simulation is significant as shown in Figure \ref{fig: gauss_tsc_ee_sim}. This can be explained by the $o(1)$ term in \eqref{eq:ber_err_exp}. Accordingly, when $L$ is increased to about 40, we see the theory and simulation curves merging as shown in Figure \ref{fig: gauss_tsc_ee_sim}. Figure \ref{fig: gauss_tsc_ee_sim2} shows that when $\rho_s$ is moderately high, smaller $L$ is required to get the performance match between theory and simulation.

From the various simulation plots in Figures \ref{fig: Pe_w_D_Gaussian_Cauchy}, \ref{fig: pe_tsc_diff_dist_L}, \ref{fig: eta_psc_diff_dist}, and \ref{fig: eta_tsc_diff_dist}, we see that the proposed scheme is robust in the sense that it works very well for a variety of sensing noise distributions including the impulsive Laplacian distribution and the Cauchy distribution which has no finite moments.

\subsection{Non-Gaussian Channel Noise} \label{sec: simnonGaussianChannel}
\begin{figure}[tb]
\begin{minipage}{1\textwidth}
\centering
\begin{center}
\includegraphics[height=9cm,width=12cm]{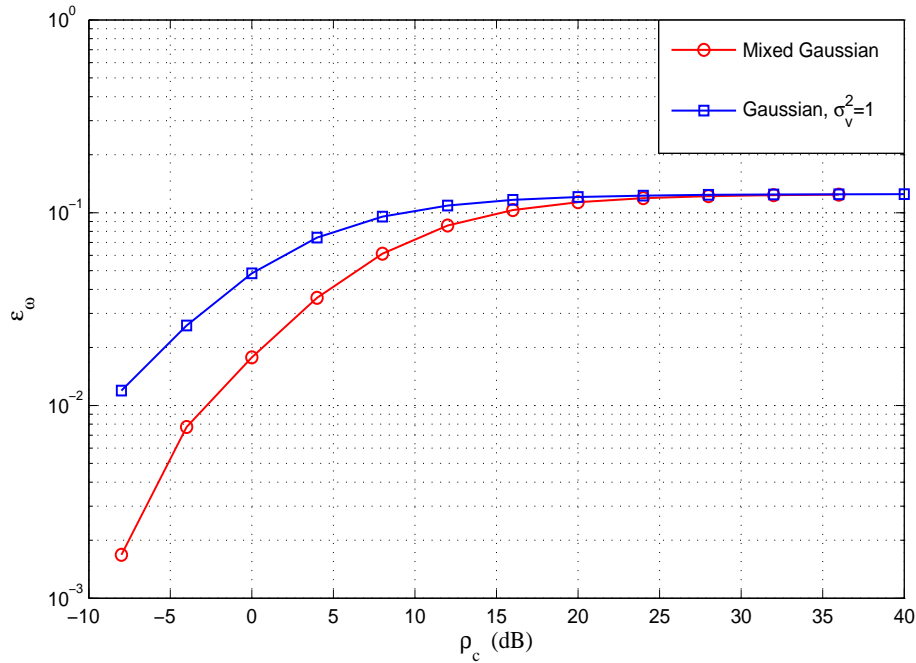}
\caption{Mixed Gaussian channel noise: $\varepsilon_{\omega}$ vs $\rho_c$: $\sigma_{v_{0}}^{2}$=0.25, $p_0=0.8$, $\sigma_{v_{1}}^{2}$=4, $p_1=0.2$, $\sigma_{v_{\rm eff}}^2=1$, $\rho_s$=0 dB.}\label{fig: Mixed_Gaussian} 
\end{center}
\end{minipage}
\end{figure}

Figure \ref{fig: Mixed_Gaussian} shows the error exponent plot for the case where $\sigma_{v_{0}}^2=0.25, p_0=0.80, \sigma_{v_{1}}^2=4, p_1=0.20$ (note that the effective channel noise variance is: $\sigma_{v_{\rm eff}}^2=p_0 \sigma_{v_{0}}^2 + p_1 \sigma_{v_{1}}^2=1$). We see that the error exponent of mixed Gaussian with $\sigma_{v_{\rm eff}}^2=1$ is worse compared to that of the Gaussian with $\sigma_{v}^2=1$ case. This is because, in the mixed Gaussian case, the error exponent is a function of the larger variance of $\sigma_{v_{1}}^2=4$. 

\begin{figure}[tb]
\begin{minipage}{1\textwidth}
\centering
\begin{center}
\includegraphics[height=9cm,width=12cm]{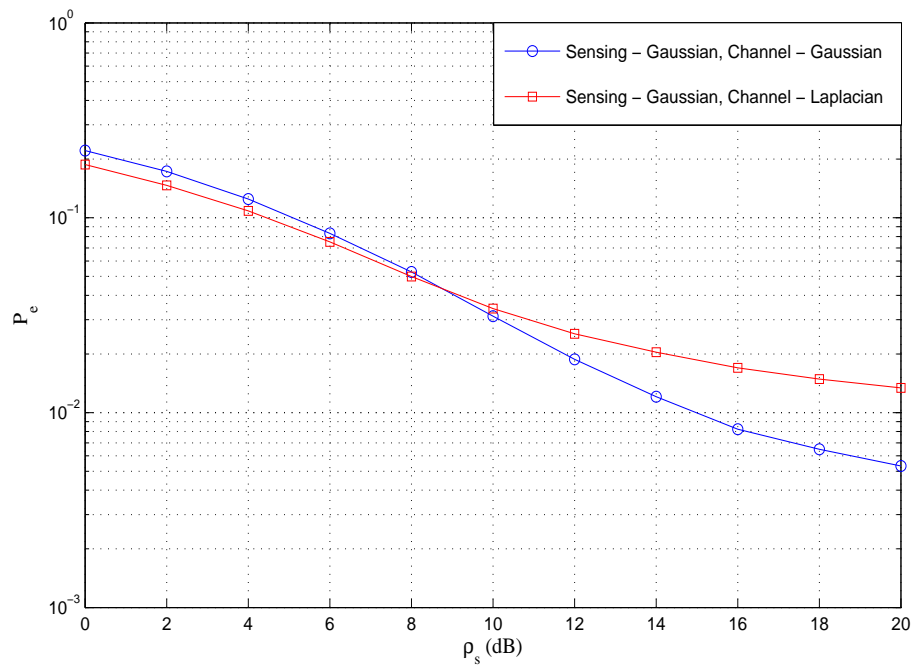}
\caption{Gaussian Sensing Noise: $P_{\rm e} $ vs $\rho_s$: $P_{\rm T}$=-12.22 dB, $L$=60}\label{fig: ch_noise_Laplacian}
\end{center}
\end{minipage}
\end{figure}

Figure \ref{fig: ch_noise_Laplacian} shows the performance of the proposed detector with Laplacian channel noise against the Gaussian channel noise when the sensing noise is Gaussian. We note that when sensing SNR $\rho_s$ is moderately high, the impulsive Laplacian channel noise is worse compared to Gaussian channel noise.

\section{Conclusions} \label{Conclusions}
A distributed detection scheme relying on constant modulus transmissions from the sensors is proposed over a Gaussian multiple access channel. The instantaneous transmit power does not depend on the random sensing noise, which is a desirable feature for low-power sensors with limited peak power capabilities. The DC of the proposed scheme is shown to depend on the characteristic function of the sensing noise and optimized with respect to $\omega$ for various sensing noise distributions. In addition to the desirable constant-power feature, the proposed detector is robust to impulsive noise, and performs well even when the moments of the sensing noise do not exist as in the case of the Cauchy distribution. Extensions to non-homogeneous sensors with non-identically distributed noise are also considered. It is shown that over Gaussian multiple access channels, the proposed detector outperforms AF, DF and MDF schemes consistently, and the MAF scheme when the sensing SNR is greater than 4 dB. The proposed detector is shown to work with the non-Gaussian channel noises as well. The error exponent is also derived for the proposed scheme and large deviation theory is used to approximate $P_{\rm e}(\omega)$ for large $L$. It is  shown that while the DC has a simpler expression for the purpose of optimizing $\omega$, the probability of error approximation based on \eqref{eq:gartner_ellis2} is shown to be an accurate indicator of detection performance for all distributions and moderate number of sensors. The effect of fading is also considered, and shown to be detrimental to the detection performance.

\section*{Appendix 1 : Gaussian Sensing Noise} \label{sec: Appendix1}
First we note that $G_{\rm G}(0)=(\alpha -1) >0$ since $\sigma_v^2 >0$ and $G_{\rm G}(\pi)= -\infty$. Since $G_{\rm G}(\beta)$ is continuous,  \eqref{gauss_first_der} has at least one solution. To show that this solution is unique, consider the first derivative:
\begin{equation}
\label{gauss_sec_der}
G^{'}_{\rm G}(\beta) = \frac{\sigma_n^2}{\theta^2} \left [ 2 \beta e^{- \frac{\sigma_n^2}{\theta^2} \beta^{2}} - 2 \alpha \left ( \frac {\beta}{2} \sec^{2}{\left (\frac {\beta}{2}\right)}  + \tan \left (\frac {\beta}{2}\right) \right) \right ].
\end{equation}
Now, using $\tan(\beta/2) \geq \beta/2$ and $\sec^{2}(\beta/2) \geq 1 + (\beta^2/4)$ for $\beta \in (0, \pi)$, we get the following upper bound:
\begin{equation}
\label{gauss_sec_der2}
G^{'}_{\rm G}(\beta) \leq  \frac{\sigma_n^2}{\theta^2} \left [2\beta e^{- \frac{\sigma_n^2}{\theta^2} \beta^{2}} - \alpha \beta \left (1+ \frac {\beta^2}{4}\right) - \alpha \beta \right ].
\end{equation}
Since $\sigma_v^2>0$ we have $\alpha>1$. Recall that $\beta \in (0, \pi)$, and the rhs of \eqref{gauss_sec_der2} is always negative. It follows that $G_{\rm G}(\beta)$ is monotonically decreasing over $\beta \in (0, \pi)$ and \eqref{gauss_first_der} has a unique solution which corresponds to the global maximum of $D(\beta/ \theta)$.

\section*{Appendix 2 : Cauchy Sensing Noise} \label{sec: Appendix2}
The first derivative of $D(\beta/ \theta)$ is given by,
\begin{equation}
\label{cauchy_first_der}
D^{'}\left (\frac {\beta}{\theta}\right) = \left [ \frac{\sin (\beta) e^{\frac{2 \gamma}{\theta} \beta}}{(\alpha e^{\frac{2 \gamma}{\theta} \beta} - 1)^2} \right ] \left [\alpha - e^{- \frac{2 \gamma} {\theta} \beta} - \frac{2 \gamma} {\theta} \alpha \tan \left (\frac {\beta}{2}\right)\right ]  .
\end{equation}
Since the first term on the rhs of \eqref{cauchy_first_der} is non-zero for $\beta \in (0, \pi)$, we need to solve
\begin{equation}
\label{cauchy_first_der1}
G_{\rm C}(\beta):= \alpha - e^{- \frac{2 \gamma} {\theta} \beta} - \frac{2 \gamma} {\theta} \alpha \tan \left (\frac {\beta}{2}\right)=0 .
\end{equation}
First we see that $G_{\rm C}(0)=(\alpha -1) >0$ and $G_{\rm C}(\pi)= -\infty$ which implies that there is at least one solution to \eqref{cauchy_first_der1} in $\beta \in (0, \pi)$ as $G_{\rm C}(\beta)$ is continuous. The second derivative of $G_{\rm C}(\beta)$ is given by
\begin{equation}
\label{cauchy_first_der2}
G^{''}_{\rm C}(\beta)= - \left [ \left ( \frac{4 \gamma^2} {\theta^2} e^{- \frac{2 \gamma} {\theta} \beta} \right ) + \frac{\gamma \alpha} {\theta}  \sec^{2} \left (\frac {\beta}{2}\right)  \tan \left (\frac {\beta}{2}\right) \right ].
\end{equation}
Clearly, $G^{''}_{\rm C}(\beta) <0$ for $\beta \in (0, \pi)$ which establishes that $G_{\rm C}(\beta)$ is concave. Therefore, \eqref{cauchy_first_der1} has a unique solution which corresponds to the global maximum of $D(\beta/ \theta)$. 

\section*{Appendix 3 : Laplacian Sensing Noise} \label{sec: Appendix3}
First we note that $G_{\rm L}(0)=(1 - (1/\alpha)) >0$ if $\sigma_v^2 >0$ and $G_{\rm L}(\pi)= -\infty$. This means that \eqref{lapl_first_der} has at least one solution. The first derivative of $G_{\rm L}(\beta)$ is given by,
\begin{equation}
\label{lapl_sec_der}  
G^{'}_{\rm L}(\beta) = \frac{ 2 b^2}{\theta^2} \left [ 2 \beta \left (1 + \frac{b^2}{\theta^2} \beta^2 \right )	- \left (\beta + \frac{b^2}{\theta^2} \beta^3 \right) \sec^{2}{\left (\frac {\beta}{2}\right)} + 2 \left (1 + 3 \frac{b^2}{\theta^2} \beta^2\right) \tan \left (\frac {\beta}{2}\right) \right ] .
\end{equation}
Now, using $\tan(\beta/2) \geq \beta/2$ and $\sec^{2}(\beta/2) \geq 1 + (\beta^2/4)$ over $\beta \in (0, \pi)$ in \eqref{lapl_sec_der} and simplifying, we get the following upper bound:
\begin{equation}
\label{lapl_sec_der2}
G^{'}_{\rm L}(\beta) \leq  - \frac{ b^2}{ 2 \theta^4} \left [ (\theta^2+8 b^2) \beta^3 + b^2 \beta^5 \right ]
\end{equation}
Clearly, for $\beta \in (0, \pi)$, the rhs of \eqref{lapl_sec_der2} is always negative which implies $G^{'}_{\rm L}(\beta)<0$. It follows that $G_{\rm L}(\beta)$ is monotonically decreasing over $\beta \in (0, \pi)$ and \eqref{lapl_sec_der2} has a unique solution which corresponds to the global maximum of $D(\beta/ \theta)$. 

\section*{Appendix 4: Proof of Theorem \ref{thm3}} \label{sec: Appendix4}
We use the G$\ddot{\rm a}$rtner-Ellis theorem from large deviation theory \cite[pp. 14]{Bucklew} to calculate the error exponent. To this end, we need to calculate $M_{\omega}(t)$ in \eqref{eq:mgf} and substitute into \eqref{eq:gartner_ellis2}.
\begin{align}
\nonumber M_{\omega}(t) &=\lim_{L \rightarrow \infty} \frac{1}{L}\log {\rm E} \lbrace \exp [t Z_L] \rbrace\\ 
\nonumber	 &=\lim_{L \rightarrow \infty} \frac{1}{L}\log {\rm E} \left \lbrace \exp \left [t \left ( \displaystyle\sum_{i=1}^{L} 2 \sin \left (\frac{\omega \theta}{2} \right) \cos \left (\omega n_{i} - \frac{\omega \theta}{2}+\frac{\pi}{2} \right ) + \sqrt{\frac{L}{P_{\rm T}}} v^{\rm T} \right ) \right ] \right \rbrace\\ 
 \label{eq:log_bessel_func}
	 &= \log {\rm E} \left \lbrace \exp \left [2 t \sin \left (\frac{\omega \theta}{2} \right) \cos \left( \omega n_{i} - \frac{\omega \theta}{2}+\frac{\pi}{2} \right ) \right ] \right \rbrace  + \left [ \frac{ t^{2}  \sigma_{v}^{2}(1-\cos (\omega\theta))}{2P_{\rm T}} \right ]
\end{align}
From \cite[pp. 376]{milton}, we have the Fourier series expansion of the periodic function $e^{p \cos(u)}$ as,
\begin{equation}
\label{eq:fourier_series_exp}
e^{p \cos(u) } = I_{0}(p) + 2 \displaystyle \sum_{k=1}^{\infty} I_{k}(p) \cos (ku)
\end{equation}
Using the equation (\ref{eq:fourier_series_exp}) in (\ref{eq:log_bessel_func}) with $p=2  t \sin( \omega \theta/2 )$ and $u=\left ( \omega n_i - \omega \theta/2 + \pi/2 \right )$ and then applying the expectation on the resulting summation, we get $M_{\omega}(t)$ as in \eqref{eq:err_m_t}.
\bibliographystyle{IEEEtran}
\bibliography{pm_bib_file}
\newpage

\end{document}